\pdfoutput=1

\documentclass[11pt,a4paper]{article}
\usepackage{pasteurlabs}

\usepackage{amsfonts}
\usepackage{nicefrac}
\usepackage{multirow}
\usepackage[capitalize,noabbrev]{cleveref}
\usepackage{orcidlink}

\newif\ifpreprint
\preprinttrue

\makeatletter
\def\fps@figure{!htbp}
\def\fps@table{!htbp}
\makeatother

\title{Mosaic: A Benchmark Suite for Differentiable Physics Solvers}

\author{Andrin Rehmann\,\orcidlink{0009-0001-6002-5096},\;
        Heiko Zimmermann\,\orcidlink{0000-0001-9190-4328},\;
        Dion Häfner\,\orcidlink{0000-0002-4465-7317}}

\date{11-2026}

\organization{\includegraphics[height=26pt]{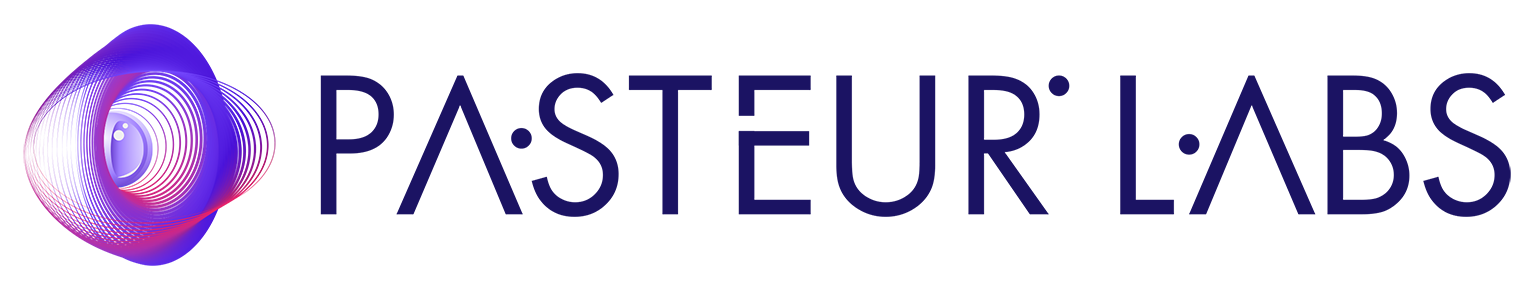}}

\affiliation{Pasteur Labs, Brooklyn, NY, USA}

\correspondence{dion.haefner@simulation.science}

\shorttitle{Mosaic: A Benchmark Suite for Differentiable Physics Solvers --- Rehmann et al., 2026}

\begin{document}

\maketitle

\begin{plabstract}
Differentiable partial differential equation (PDE) solvers underpin
solver-in-the-loop ML training, gradient-based optimal control, and
inverse problems, yet the practical cost of obtaining correct, usable
gradients from a given solver on a given problem is largely
undocumented. Integration effort, computational cost, memory, gradient accuracy,
and numerical stability vary widely across solvers and are
discoverable only by trial and error.
We introduce \textit{Mosaic}, an extensible benchmarking framework for
differentiable PDE solvers that standardizes access to solver gradients.
Each solver is packaged as a containerized component
(\textit{Tesseract}) exposing a uniform gradient API regardless of
language or automatic differentiation (AD) strategy, enabling
researchers to evaluate, compare, and build on non-trivial physical solvers.
Our evaluation of 14 solvers across fluid dynamics, structural
mechanics, and heat transfer demonstrates that the benchmark surfaces
practically relevant differences: order-of-magnitude variation in
computational cost and memory, alongside setup
incompatibilities and instabilities that rule out solvers for specific
tasks.
Despite this variation, all solvers that
produce gradients converge to similar optima on our tasks, indicating that the
practical barriers are memory limits, numerical stability, and setup
compatibility rather than gradient accuracy alone.
Mosaic is open-source and available at \href{https://github.com/pasteurlabs/mosaic}{github.com/pasteurlabs/mosaic}.

\end{plabstract}

\begin{figure}[h!]
  \centering
  \includegraphics[width=\linewidth]{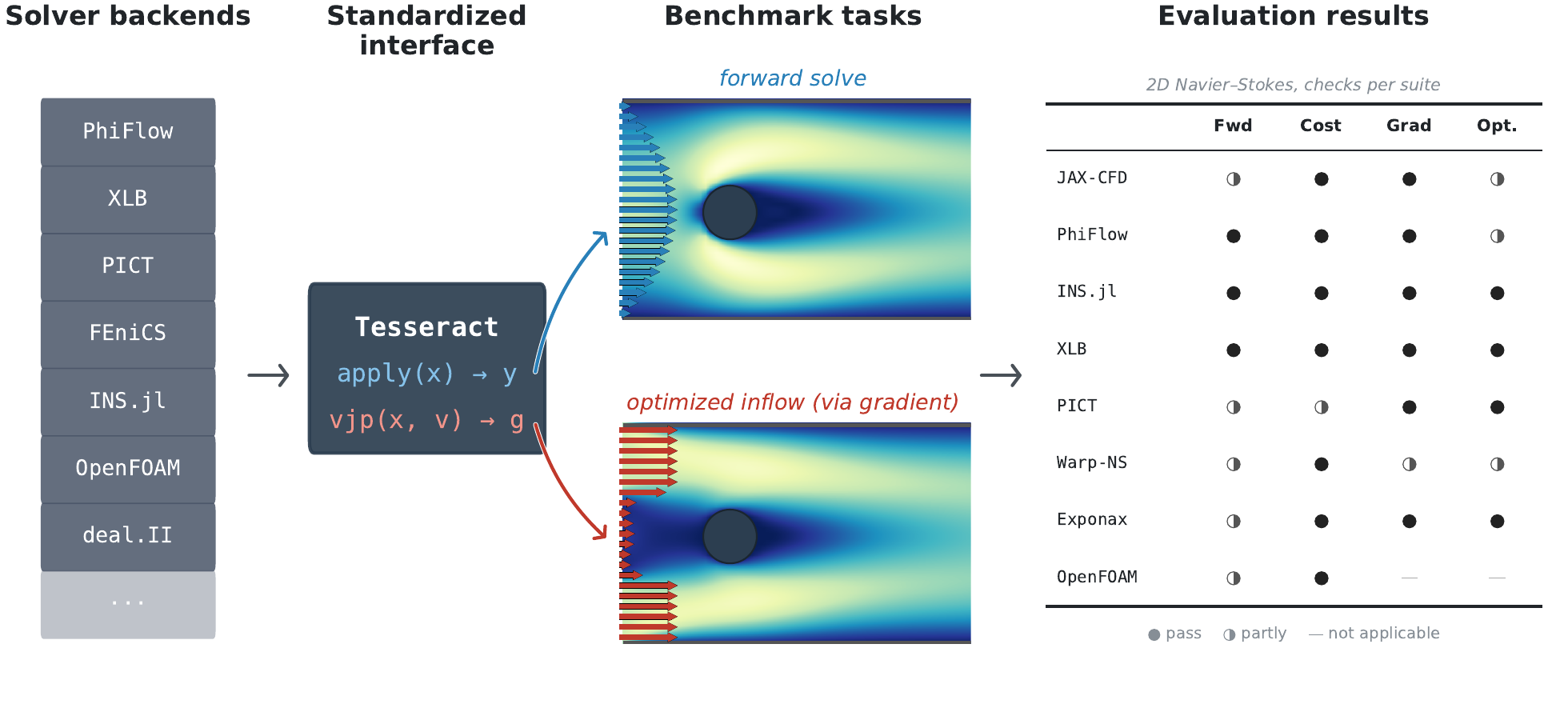}
  \caption{Overview of Mosaic. Diverse solvers are
    wrapped behind a uniform containerized interface, enabling cross-solver comparison on shared benchmark tasks involving different physical domains.}
  \label{fig:teaser}
\end{figure}

\section{Introduction}
\label{sec:intro}

Differentiable solvers enable topology
optimization~\cite{bendsoe2003topology}, aerodynamic shape
optimization~\cite{jameson1988aerodynamic}, optimal
control~\cite{bewley2001dns}, and solver-in-the-loop ML
training~\cite{kochkov2021machine, um2020sol} by backpropagating
through a numerical simulation's forward pass.
Differentiating through ODE integrators~\cite{chen2018neural,rackauckas2021universal}
and renderers~\cite{kato2020differentiable} is now routine, but PDE
solvers are more complex, less standardized, and their gradient
properties far less understood.

A growing number of physics solvers advertise gradient support,
spanning multiple languages, AD frameworks, and numerical methods.
Yet it is hard to know in advance what reliable gradients from any of
them will cost. Solvers range from turnkey applications to
composable frameworks and low-level kernel toolkits, each demanding
different expertise and integration effort. Runtime and memory overhead,
gradient accuracy, and stability over long rollouts have not been
systematically measured. Solver-specific constraints therefore surface
only after a practitioner has committed to a particular solver: some
solvers support only periodic domains or lack the boundary types a task
needs, some become unstable or consume excessive resources over long rollouts,
and some lose accuracy in settings their own tests do not cover.

Learned surrogates and physics foundation models offer an alternative
path, with limitations of their own (\cref{sec:related}). Meanwhile,
decades of engineering have produced solvers that practitioners
trust for forward simulation. What is missing is not the solvers
themselves, which exist in abundance, but a standardized way to
access, compare, and trust their gradients.

We introduce \textit{Mosaic}, an extensible benchmark suite that
provides a uniform interface to a diverse set of differentiable solvers
across multiple physical domains, with benchmark tasks that are designed
to systematically isolate issues with solver gradients. An initial set of
tasks focuses on standalone optimization
problems without learned components, where finite differences provide
ground truth and the solver's gradient mechanism is the only variable
under test. Mosaic enables researchers to evaluate which solver fits their problem,
build ML and optimization pipelines on top of the
uniform interface, and stress-test failure modes before
committing to a particular stack. Our contributions are:
\begin{enumerate}\itemsep2pt
  \item A \textbf{benchmark suite and containerized solver interface}
    spanning 14 solvers across four physical domains (heat transfer,
    structural mechanics, 2D and 3D Navier--Stokes;
    \cref{tab:domains}). Each solver is wrapped behind a uniform
    forward/gradient API regardless of language or AD strategy, so
    researchers can run and compare real solvers without
    solver-specific expertise.
  \item A \textbf{standardized evaluation protocol} measuring gradient
    accuracy, computational cost, Jacobian conditioning, setup
    compatibility, and end-to-end optimization convergence.
  \item \textbf{Empirical findings} showing that, on our tasks, all solvers
    that produce gradients reach comparable optima, and that solvers are ruled
    out instead by setup incompatibilities, instabilities over long
    rollouts, and the memory cost of their gradients.
  \item An \textbf{open-source release} of all solver interfaces, evaluation
    code, and templates, extensible to new solvers and domains, with
    CI-based re-evaluation infrastructure so that published results stay
    current as solvers improve.
\end{enumerate}

\section{Background and related work}
\label{sec:related}

\paragraph{Adjoint methods and differentiable physics}
Adjoint methods for PDE-constrained optimization date back to
Lions~\cite{lions1971optimal}, and were brought to fluid design by
Pironneau~\cite{pironneau1974optimum} and
Jameson~\cite{jameson1988aerodynamic}, with discrete adjoint
formulations following in~\cite{giles2000introduction,nadarajah2000studies}.
Modern AD frameworks now differentiate solver code automatically:
JAX source transformation~\cite{bradbury2018jax}, PyTorch operator
overloading~\cite{paszke2019pytorch}, Julia source-to-source AD via
Zygote~\cite{innes2018don}, LLVM-IR-level AD via
Enzyme~\cite{moses2020enzyme}, NVIDIA
Warp kernel-level AD~\cite{macklin2024warp}, and tape-based discrete
adjoints in dolfin-adjoint and
firedrake-adjoint~\cite{farrell2013automated,mitusch2019dolfinadjoint}.
Hand-derived adjoints are fast but tied to a fixed set of objectives,
requiring new derivations whenever the control variable or loss function
changes. Algorithmic AD removes this constraint at the price of higher
memory or computational cost, a tradeoff whose magnitude has remained
largely anecdotal. Mosaic measures this tradeoff empirically across solver families.

\paragraph{Learned surrogates and physics foundation models}
Neural operator surrogates~\cite{kovachki2023neural,li2021fourier,lu2021learning},
PINNs~\cite{raissi2019physics}, and physics foundation
models~\cite{pathak2022fourcastnet,bodnar2024aurora,price2024gencast,hao2024dpot,herde2024poseidon}
are complementary to differentiable solvers but introduce approximation
error, degrade outside the training distribution, require solver-generated
training data in the first place, and so far cover only a few well-studied
PDE families. Mosaic benchmarks the solver layer directly.

\paragraph{Existing benchmarks and evaluation efforts}
Several benchmarks evaluate PDE solvers or learned surrogates, but none
systematically measures gradient properties across solver backends.
APEBench~\cite{koehler2024apebench} benchmarks neural PDE emulators on
forward accuracy without evaluating solver gradients.
PDEBench~\cite{takamoto2022pdebench} provides datasets of PDE solutions
for surrogate training but includes no gradient evaluation.
NeuralFluid~\cite{fei2024neuralfluid} pairs a differentiable
Navier--Stokes solver with a benchmark of design and control tasks and
compares its forward and backward runtime against PhiFlow, yet
does not assess gradient accuracy, conditioning, or setup compatibility across backends.
List et al.~\cite{list2024differentiability} analyze gradient behavior in
unrolled training, focusing on convergence dynamics rather than
cross-solver comparison.
The PhiFlow benchmark~\cite{holl2024phiflow} compares AD backends within
a single framework, not across independent solver implementations.
preCICE~\cite{bungartz2016precice} standardizes solver coupling but
provides no gradient interface.

The closest analogue in design philosophy is OpenAI Gym~\cite{brockman2016openai}, which standardized access to RL
environments behind a uniform \texttt{step}/\texttt{reset} interface.
Gym ships no RL algorithms, yet the standardized interface enabled a
wealth of downstream research by lowering the barrier to entry.
Mosaic follows the same model for differentiable solvers: solver
backends are simulators behind a standardized gradient interface, and
the benchmark ships no ML components itself but provides the
infrastructure that researchers and practitioners can build on.
\Cref{sec:sil} reports one such study.

\section{The Mosaic benchmark suite}
\label{sec:mosaic}

Mosaic organizes evaluation around three concepts. A \emph{physical
setup} is a concrete problem instance: a specific geometry, boundary
conditions, and operating regime within a physical domain (e.g.,
incompressible fluids). A \emph{benchmark task} is a fixed optimization
problem on a specific setup: a control variable, an objective, and a
reference solution. A \emph{solver backend} is
the software that evaluates the forward map and (optionally) its
gradient. Every solver that supports a
task's physical domain can be compared on it.

\subsection{Physical domains and benchmark tasks}
\label{sec:domains}

\begin{figure}[t]
  \centering
  \begin{subfigure}[t]{0.48\linewidth}
    \centering
    \includegraphics[width=0.95\linewidth]{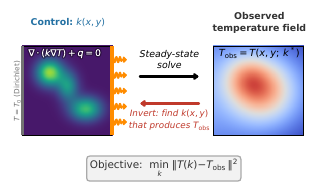}
    \caption{\textbf{H}\,: Conductivity inversion in steady heat conduction.}
    \label{fig:domain_h}
  \end{subfigure}\hfill
  \begin{subfigure}[t]{0.48\linewidth}
    \centering
    \includegraphics[width=0.95\linewidth]{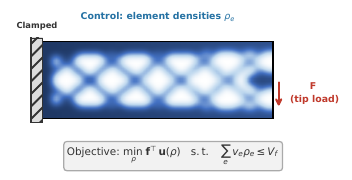}
    \caption{\textbf{S}\,: Compliance minimization of a cantilever (SIMP).}
    \label{fig:domain_s}
  \end{subfigure}\\[0.2em]
  \begin{subfigure}[t]{0.48\linewidth}
    \centering
    \includegraphics[width=0.95\linewidth]{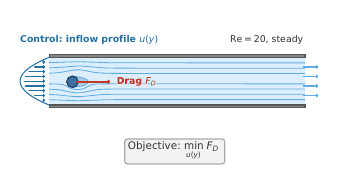}
    \caption{\textbf{F2}\,: Inflow optimization for a 2D cylinder.}
    \label{fig:domain_f2}
  \end{subfigure}\hfill
  \begin{subfigure}[t]{0.48\linewidth}
    \centering
    \includegraphics[width=.85\linewidth]{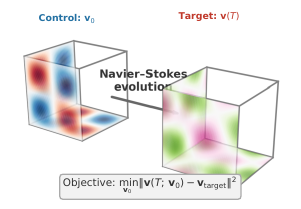}
    \caption{\textbf{F3}\,: Initial condition (IC) recovery in a 3D periodic box.}
    \label{fig:domain_f3}
  \end{subfigure}
  \caption{Benchmark tasks across the four physical domains in Mosaic.
    Each panel shows the control variable, the physical process,
    and the optimization objective.}
  \label{fig:domains}
\end{figure}

Benchmark tasks are representative optimization problems (inflow
optimization, topology optimization, inverse problems) that require
gradients for efficient solution, are supported by multiple backends
for cross-solver comparison, and contain no learned components so solver
gradients can be validated in isolation. Tasks are intentionally simpler
than state-of-the-art problems, as complexity in the physics could
confound the gradient properties we aim to isolate. This allows us to
attribute failures to the solver's gradient computation rather than to an
ill-posed optimization problem. The heat domain (H) additionally serves
as a negative control. Its forward problem is smooth and elliptic, so
any gradient failure on H is attributable to an AD bug rather than to
chaotic dynamics.
\Cref{tab:domains,fig:domains} summarize the four domains, and
\cref{app:effort} gives their full specifications.

\begin{table}[t]
  \centering\small
  \caption{Physical domains and optimization tasks. The backend count
    includes forward-only references; per-solver details are in
    \cref{tab:solvers}.}
  \label{tab:domains}
  \begin{tabularx}{\linewidth}{c>{\hsize=0.95\hsize}X>{\hsize=1.05\hsize}Xlll}
    \toprule
    \textbf{ID} & \textbf{Domain}
      & \multicolumn{2}{c}{\textbf{Optimization task}} & \textbf{Backends} \\
    \cmidrule(lr){3-4}
    & & \textit{Task} & \textit{Control dim.} & \\
    \midrule
    H & Heat transfer (slab) & Conductivity inversion
      & 128 & 5 \\
    & \multicolumn{2}{l}{\textit{%
      Negative control: smooth physics to isolate gradient bugs}} & & \\[5pt]
    S & Structural mech.\ (cantilever) & Compliance minimization (SIMP)
      & 256 & 6 \\
    & \multicolumn{2}{l}{\textit{%
      Well-studied topology optimization with known optimum}} & & \\[5pt]
    F2 & Incompr.\ fluids (2D cylinder) & Inflow optimization for drag min.
      & 41 & 8 \\
    & \multicolumn{2}{l}{\textit{%
      Anchor domain: standard channel benchmark, plus periodic checks}} & & \\[5pt]
    F3 & 3D Navier--Stokes (periodic) & Initial condition recovery
      & 12\,288 & 8 \\
    & \multicolumn{2}{l}{\textit{%
      Scale test: large states and long rollouts in a periodic box}} & & \\
    \bottomrule
  \end{tabularx}
\end{table}

\subsection{Design principles}
\label{sec:scope}

Mosaic consists of three artifacts: a containerized solver interface,
an evaluation protocol, and a library of benchmark task templates.
Containerization is handled by Tesseract~\cite{haefner2025tesseract},
an existing open-source runtime for packaging scientific codes. Mosaic
contributes the evaluation methodology, benchmark tasks, and empirical
results on top of it. The four domains and 14 solver backends described
in this paper are its initial instantiation.

\paragraph{Solver interface and extensibility}
Each solver is wrapped as a containerized component exposing two entrypoints,
\texttt{apply(inputs)\,$\rightarrow$\,outputs} and
\texttt{vector\_jacobian\_product(inputs, cotangents)\,$\rightarrow$\,vjp}.
The container isolates all dependencies (language, libraries, GPU
runtime), so any of the 14 backends can be swapped by changing a single
configuration string.
Forward-only solvers
(OpenFOAM, deal.II) implement \texttt{apply} only and serve as forward
reference baselines.
Because solver APIs differ too widely for a single generic wrapper, a
\emph{task-specific interface schema} (inputs, outputs, shapes, and units) is
defined for each benchmark task, and multiple solvers implement it. To add a
backend, a contributor packages the solver as a Tesseract that implements
the task schema, and the evaluation suite re-runs automatically on every
proposed change.

Each benchmark task acts as a validated template with known-good settings,
so users can modify the physical setup (geometry, Reynolds number, boundary conditions) while
inheriting the evaluation protocol and finite-difference reference.
\Cref{app:software} describes how to contribute, and \cref{app:cost}
quantifies the overhead of containerization.

\subsection{Evaluation protocol}
\label{sec:protocol}

All metrics are computed uniformly across solvers. The protocol is
solver-agnostic: it operates on the containerized interface (forward and
VJP) and never inspects solver internals.

\begin{description}\itemsep2pt
  \item[Setup compatibility.] For each solver--task pair, we record
    whether the solver produces a gradient, fails numerically, or is
    structurally unable to represent the task (\cref{sec:optimization}).
  \item[Forward accuracy.] We compare accuracy across resolutions against a reference
    (a cross-solver consensus, or a converged spectral run for F3 and
    Kolmogorov flow) and verify adherence to physical laws where applicable
    (e.g., solenoidal velocity). Where an analytical solution exists (F2),
    precision is additionally measured against it.
  \item[Gradient accuracy.] Central finite differences through each
    solver's own forward pass serve as the reference. We
    evaluate directional derivatives along $K$ random perturbation
    vectors ($K{=}6$--$20$) and report cosine similarity and relative
    error for all problems. Besides the channel, F2 includes periodic
    setups that provide two further checks. For the decaying Taylor--Green
    vortex, the closed-form solution gives exact gradients with respect to
    the initial field and $\nu$ (\cref{tab:gradchecks}). On forced
    Kolmogorov flow, a common testbed for learned flow
    correction~\cite{kochkov2021machine}, we check the gradient through 16
    chained calls, as in a training loop (\cref{app:kolmogorov}). For the
    two Navier--Stokes problems (F2, F3), we also compute the full
    Jacobian and inspect its singular value spectrum.
  \item[Performance.] We measure primal and VJP wall-clock time, their
    ratio, and peak GPU (device-wide) and host memory on each solver's intended
    hardware target across various problem sizes (grid resolutions or cell counts).
    The VJP time is the cost of one gradient from scratch, a forward plus an
    adjoint pass.
    All timings are medians of 3 trials (10 on the mesh domains) after one
    warmup call to pre-populate JIT caches and stabilize memory allocation.
  \item[Optimization convergence.] We run fixed optimizers with each
    solver's gradients on each benchmark task for a fixed iteration budget,
    and compare the convergence curves and final objectives across solvers.
    \Cref{app:recovery} lists the budgets and optimizer settings, and the
    pass thresholds behind the \textit{Conv.} column of \cref{tab:merged}.
\end{description}

\subsection{Included solver backends}
\label{sec:solvers}

The solver set (\cref{tab:solvers}) is chosen to maximize diversity
along three axes: \emph{AD strategy} (autodiff, tape-based discrete
adjoints, hand-derived/kernel adjoints), \emph{numerical method} (projection-based, lattice Boltzmann,
finite-volume, spectral, finite elements with both direct and iterative
linear solvers), and \emph{solver abstraction level}.
These range from self-contained \emph{applications} (OpenFOAM),
through \emph{frameworks} that handle discretization and AD (FEniCS,
PhiFlow, JAX-CFD), to \emph{kernel toolkits} (NVIDIA Warp) where the
Mosaic wrapper \emph{is} the solver. Abstraction level determines
available gradient strategies and per-setup effort.

\begin{table}[t]
  \centering\small
  \caption{Solver backends included in the current release.
    Domain IDs match \cref{tab:domains}.}
  \label{tab:solvers}
  {\setlength{\tabcolsep}{4pt}
  \begin{tabular}{@{}llllll c@{}}
    \toprule
    \textbf{Solver} & \textbf{\shortstack[l]{Language /\\AD framework}}
      & \textbf{Strategy}
      & \textbf{\shortstack[l]{Discret-\\ization}} & \textbf{Numerics}
      & \textbf{GPU} & \textbf{Domains} \\
    \midrule
    JAX-CFD~\cite{kochkov2021machine}
      & Py/JAX & AD & FD & Projection, RK4 & $\checkmark$ & F2, F3 \\
    PhiFlow~\cite{holl2024phiflow}
      & Py/JAX & AD-S & FD & RK4/semi-Lagr. & $\checkmark$ & F2, F3 \\
    INS.jl~\cite{agdestein2024ins}
      & Julia/Zygote & AD & FV & Projection, RK4 & $\checkmark$ & F2, F3 \\
    XLB~\cite{ataei2024xlb}
      & Py/JAX & AD-C & LBM & BGK, float64 & $\checkmark$ & F2, F3 \\
    PICT~\cite{franz2025pict}
      & Py/PyTorch & AD-S & FV & PISO, BDF1 & $\checkmark$ & F2, F3 \\
    Warp-NS~\cite{macklin2024warp}
      & Py/Warp & AD/AD-S & FD & Projection, Euler & $\checkmark$ & F2, F3 \\
    Exponax~\cite{koehler2024apebench}
      & Py/JAX & AD & Spectral & ETDRK2 & $\checkmark$ & F2, F3 \\
    OpenFOAM~\cite{weller1998tensorial}
      & C++ & $-$ & FV & PISO/PIMPLE, Euler & & F2, F3 \\
    \midrule\midrule
    FEniCS~\cite{alnaes2015fenics}
      & Py & VA & FE & Direct (MUMPS) & & H, S \\
    Firedrake~\cite{rathgeber2016firedrake}
      & Py & VA & FE & Direct (MUMPS) & & H, S \\
    JAX-FEM~\cite{xue2023jaxfem}
      & Py/JAX & AD-S & FE & BiCGSTAB+Jacobi & $\checkmark$ & H, S \\
    TopOpt.jl~\cite{huang2021topoptjl}
      & Julia & HA & FE & Direct (CHOLMOD) & & S \\
    torch-fem~\cite{meyer2024torchfem}
      & Py/PyTorch & AD-S & FE & CG+Jacobi & $\checkmark$ & H, S \\
    deal.II~\cite{bangerth2007dealii,arndt2023dealii}
      & C++ & $-$ & FE & Direct (S), CG+AMG (H) & & H, S \\
    \bottomrule
  \end{tabular}}
  \par\vspace{4pt}
  \begin{minipage}{\linewidth}\footnotesize\raggedright
    \textit{Strategy:}
    AD\,=\,reverse-mode AD storing every step;
    AD-C\,=\,AD with checkpointing;
    AD-S\,=\,AD with adjoint linear solves;
    VA\,=\,variational adjoint (pyadjoint);
    HA\,=\,hand-derived adjoint;
    $-$\,=\,forward-only;
    a slash separates periodic box and channel.
    \textit{Discr.:}
    FD\,=\,finite differences;
    FV\,=\,finite volume;
    FE\,=\,finite elements;
    LBM\,=\,lattice Boltzmann method.
    \textit{Numerics:}
    PISO\,=\,pressure-implicit with splitting of operators;
    PIMPLE\,=\,merged PISO-SIMPLE;
    BDF1\,=\,1st-order backward differentiation;
    ETDRK\,=\,exponential time-differencing Runge--Kutta;
    BGK\,=\,Bhatnagar--Gross--Krook.
  \end{minipage}
\end{table}

\paragraph{Inclusion and exclusion criteria}
A differentiable solver is included if it provides a reverse-mode VJP for
arbitrary cotangents at the cost of a few forward solves, independent of
the number of inputs. This rules out, for example, solvers that offer only
forward-mode AD, whose gradient costs one pass per input dimension, as
finite differences do. Established forward-only solvers are included as
reference baselines for forward accuracy. Excluded solvers and reasons are
documented in \cref{app:solvers}.

\section{Experiments}
\label{sec:experiments}

The following experiments check that Mosaic's evaluation protocol
distinguishes solvers in ways that matter in practice, and they provide an
initial empirical characterization to guide solver selection.
All experiments ran on an Azure \texttt{Standard\_NC24ads\_A100\_v4} instance
(one NVIDIA A100 80\,GB PCIe, 24 vCPUs of an AMD EPYC 7V13).
Results reflect the solver configurations of the pinned
release, and reproduction instructions are in \cref{app:software}.

\subsection{Benchmark summary and implementation effort}
\label{sec:baselines}

\Cref{tab:merged} summarizes benchmark results and implementation
effort across all solvers, and the subsections below discuss each metric.
Physical-accuracy validation is in \cref{app:physical_accuracy}, and the
reasons for setup incompatibilities are in \cref{app:exclusions}.

Implementation effort varies substantially among differentiable solvers.
We report bespoke lines of code (LOC) as a coarse proxy for integration
complexity, an imperfect metric but one that captures order-of-magnitude
differences reproducibly (\cref{app:effort}).
AD framework support does not guarantee turnkey gradients. PhiFlow, JAX-CFD,
and Exponax need at most 20 lines of gradient code, whereas INS.jl, PICT, and
XLB, which also differentiate through an AD framework, need 91 to 228.
The live version of Mosaic also records every departure from each solver's
default usage, together with how it works, whether it is required, and the
evidence for it, as guidance for practitioners integrating that solver.

\begin{table}
  \centering\footnotesize
  \setlength{\tabcolsep}{2.5pt}
  \caption{Benchmark results, gradient support, and integration effort per solver.
    Wall-clock times are medians of 3 (F2, F3) or 10 (S, H) timed trials
    after 1 warmup. Full specifications in \cref{app:effort}; initial conditions in
    \cref{app:ics}. Legend below table.}
  \label{tab:merged}
  \vspace{2pt}
  \begin{tabular}{@{}l r r c c c c | ccc | r r r@{}}
    \toprule
    \textbf{Solver}
      & \multicolumn{2}{c}{\textbf{Wall time}}
      & \multicolumn{3}{c}{\textbf{Error}}
      & \textbf{Conv.}
      & \multicolumn{3}{c|}{\textbf{Grad.\ support}}
      & \multicolumn{3}{c@{}}{\textbf{Integ.\ effort}} \\
    \cmidrule(lr){2-3}\cmidrule(lr){4-6}\cmidrule(lr){8-10}\cmidrule(lr){11-13}
    & \textit{Fwd} & \textit{VJP} & \textit{Ref.} & \textit{An.} & \textit{FD} &
      & \multicolumn{3}{c|}{\textit{input variables}}
      & \textit{IO} & \textit{Num.} & \textit{Conf.} \\
    \midrule
    \multicolumn{7}{@{}l}{\textit{Navier--Stokes grid (F2), wall time in ms}}
      & \textit{$v_0$} & \textit{$\nu$} & \textit{inflow} & & & \\
    \cmidrule(lr){8-10}
      Exponax~\cite{koehler2024apebench} & 26 & 46 & \textbf{3.3e-3} & \textbf{2.5e-6} & \textbf{4.1e-6} & $\bot$ & $\bullet$ & $\bullet$ & $\bot$ & 41 & 194 & \textbf{0} \\
      INS.jl~\cite{agdestein2024ins} & 160$^\dagger$ & 740 & 3.7e-3 & 2.4e-3 & 1.6e-5 & $\checkmark$ & $\bullet$ & $\bullet$ & $\bullet$ & 143 & 736 & 111 \\
      JAX-CFD~\cite{kochkov2021machine} & 20 & 40 & 3.7e-3 & 2.4e-3 & 7.1e-6 & $\bot$ & $\bullet$ & $\bullet$ & $\bot$ & 9 & 166 & \textbf{0} \\
      PhiFlow~\cite{holl2024phiflow} & 280 & 1500 & 3.7e-3 & 2.4e-3 & 1.0e-5 & $\checkmark$ & $\circ$ & $\circ$ & $\circ$ & 28 & 368 & \textbf{0} \\
      PICT~\cite{franz2025pict} & 1700 & 3900 & \textbf{3.3e-3} & 4.6e-3 & 4.6e-5 & $\checkmark$ & $\circ$ & $\circ$ & $\circ$ & 53 & 452 & 400 \\
      Warp-NS~\cite{macklin2024warp} & 83 & 320 & 3.6e-3 & 6.0e-4 & 3.7e-5 & $\checkmark$ & $\bullet$ & $\bullet$ & $\circ$ & \textbf{0} & 2405 & 14 \\
      XLB~\cite{ataei2024xlb} & \textbf{7.0} & \textbf{15} & 3.6e-3 & 6.1e-4 & 1.1e-5 & $\checkmark$ & $\bullet$ & $\bullet$ & $\bullet$ & 11 & 481 & 51 \\
      OpenFOAM~\cite{weller1998tensorial} & 1200 & $-$ & 3.6e-3 & 1.9e-3 & $-$ & $-$ & $-$ & $-$ & $-$ & 259 & \textbf{100} & 508 \\
    \midrule
    \multicolumn{7}{@{}l}{\textit{Navier--Stokes grid (F3), wall time in ms}}
      & \textit{$v_0$} &  &  & & & \\
    \cmidrule(lr){8-8}
      Exponax~\cite{koehler2024apebench} & 18 & 31 & \textbf{7.8e-4} & $-$ & 1.2e-4 & $\checkmark$ & $\bullet$ &  &  & 41 & 194 & \textbf{0} \\
      INS.jl~\cite{agdestein2024ins} & 140$^\dagger$ & 560 & 4.2e-2 & $-$ & 2.9e-5 & $\checkmark$ & $\bullet$ &  &  & 143 & 736 & 111 \\
      JAX-CFD~\cite{kochkov2021machine} & 19 & 37 & 4.2e-2 & $-$ & 2.0e-5 & $\checkmark$ & $\bullet$ &  &  & 9 & 166 & \textbf{0} \\
      PhiFlow~\cite{holl2024phiflow} & 160 & 370 & 4.0e-2 & $-$ & 1.8e-5 & $\checkmark$ & $\circ$ &  &  & 28 & 368 & \textbf{0} \\
      PICT~\cite{franz2025pict} & 550 & 1600 & 8.2e-3 & $-$ & 2.8e-5 & $\checkmark$ & $\circ$ &  &  & 53 & 452 & 400 \\
      Warp-NS~\cite{macklin2024warp} & 100 & 350 & 3.5e-3 & $-$ & 1.4e-5 & $\checkmark$ & $\bullet$ &  &  & \textbf{0} & 2405 & 14 \\
      XLB~\cite{ataei2024xlb} & \textbf{12} & \textbf{30} & 8.5e-3 & $-$ & \textbf{4.1e-6} & $\checkmark$ & $\bullet$ &  &  & 11 & 481 & 51 \\
      OpenFOAM~\cite{weller1998tensorial} & 3600 & $-$ & 2.2e-2 & $-$ & $-$ & $-$ & $-$ &  &  & 259 & \textbf{100} & 508 \\
    \midrule
    \multicolumn{7}{@{}l}{\textit{Structural mechanics (S), wall time in s}}
      & \textit{$\rho$} &  &  & & & \\
    \cmidrule(lr){8-8}
      FEniCS~\cite{farrell2013automated} & 4.2 & 8.7 & 0 & $-$ & 2.1e-5 & $\checkmark$ & $\circ$ &  &  & \textbf{0} & 334 & 63 \\
      Firedrake~\cite{rathgeber2016firedrake} & 1.8 & 3.9 & 0 & $-$ & 2.1e-5 & $\checkmark$ & $\circ$ &  &  & 39 & 263 & 137 \\
      JAX-FEM~\cite{xue2023jaxfem} & 1.5$^\dagger$ & 2.6$^\dagger$ & 0 & $-$ & 2.1e-5 & $\checkmark$ & $\circ$ &  &  & \textbf{0} & 203 & 109 \\
      TopOpt.jl~\cite{huang2021topoptjl} & 0.58$^\dagger$ & 0.62$^\dagger$ & 0 & $-$ & 2.1e-5 & $\checkmark$ & $\circ$ &  &  & 125 & 110 & \textbf{38} \\
      torch-fem~\cite{meyer2024torchfem} & \textbf{0.27} & \textbf{0.44} & 0 & $-$ & 2.1e-5 & $\checkmark$ & $\circ$ &  &  & \textbf{0} & \textbf{89} & 56 \\
      deal.II~\cite{bangerth2007dealii} & 2.4 & $-$ & 0 & $-$ & $-$ & $-$ & $-$ &  &  & 408 & 369 & 147 \\
    \midrule
    \multicolumn{7}{@{}l}{\textit{Heat transfer (H), wall time in s}}
      & \textit{$\rho$} & \textit{src} &  & & & \\
    \cmidrule(lr){8-9}
      FEniCS~\cite{farrell2013automated} & 1.0 & 2.3 & 6.1e-8 & $-$ & 1.8e-5 & $\checkmark$ & $\circ$ & $\circ$ &  & \textbf{0} & 507 & \textbf{61} \\
      Firedrake~\cite{rathgeber2016firedrake} & 0.37 & 0.74 & 6.1e-8 & $-$ & 1.8e-5 & $\checkmark$ & $\circ$ & $\circ$ &  & \textbf{0} & 305 & 102 \\
      JAX-FEM~\cite{xue2023jaxfem} & 0.38 & 0.77 & 6.1e-8 & $-$ & 1.8e-5 & $\checkmark$ & $\circ$ & $\circ$ &  & \textbf{0} & 275 & 100 \\
      torch-fem~\cite{meyer2024torchfem} & \textbf{0.14} & \textbf{0.20} & 6.1e-8 & $-$ & 1.8e-5 & $\checkmark$ & $\circ$ & $\circ$ &  & \textbf{0} & \textbf{220} & 70 \\
      deal.II~\cite{bangerth2007dealii} & 1.5 & $-$ & 6.1e-8 & $-$ & $-$ & $-$ & $-$ & $-$ &  & 401 & 284 & 77 \\
    \bottomrule
  \end{tabular}
  \par\vspace{2pt}
  \begin{minipage}{\linewidth}\scriptsize\raggedright
    \textit{Wall time} at $N=64$ (F2), $N=32$ (F3),
    $n_x=128$ (S), and $n_x=256$ (H);
    $^\dagger$\,=\,the trials vary by more than 15\% (\cref{tab:app:variance}).
    \textit{Error} (relative): Ref.\,=\,vs.\ reference at $\nu=0.01$ (F2: solver consensus,
    multimode IC; F3: Exponax at $N=64$), or the largest over the density sweep vs.\ the solver
    consensus (S, H), where the solvers agree to float32 round-off; An.\,=\,vs.\ analytical Taylor--Green solution at $\nu=0.01$;
    FD\,=\,vs.\ central finite differences (median over directions, best~$\varepsilon$).
    \textit{Conv.}\ (F2 drag, F3 L-BFGS+proj IC recovery, S topology opt., H conductivity recovery):
    $\checkmark$\,=\,passes the checks of \cref{app:recovery}; $\triangle$\,=\,flagged; $\times$\,=\,failed.
    \textit{Grad.\ support}: $\bullet$\,=\,AD through the solver; $\circ$\,=\,adjoint or
    hand-written VJP for at least part of it; $\times$\,=\,incorrect; $?$\,=\,not obtained;
    $\bot$\,=\,excluded (\cref{app:exclusions}); $-$\,=\,not applicable.
    Inflow support is shown only by a converged drag optimization.
    \textit{Integ.\ effort} (bespoke LOC): IO\,=\,input/output marshalling;
    Num.\,=\,numerics \& gradient code; Conf.\,=\,solver configuration.
  \end{minipage}
\end{table}

\subsection{Gradient quality}
\label{sec:quality}

We compare each solver's gradients against central finite differences at a
solver-specific optimal perturbation size~$\varepsilon$ and, on the periodic
F2 setup, against exact Taylor--Green gradients (\cref{tab:gradchecks}).

\begin{table}[t]
  \centering\small
  \caption{Relative gradient errors on F2 at $\nu=0.01$. FD: initial-condition VJP
    against central finite differences. Analytic, $\mathrm{d}J/\mathrm{d}\nu$:
    initial-condition and viscosity gradients against the exact Taylor--Green gradients.}
  \label{tab:gradchecks}
  \begin{tabular}{@{}l c c c@{}}
    \toprule
    \textbf{Solver} & \textbf{FD} & \textbf{Analytic} & $\mathrm{d}J/\mathrm{d}\nu$ \\
    \midrule
      Exponax & \textbf{4.1e-6} & \textbf{1.3e-5} & \textbf{8.9e-6} \\
      INS.jl & 1.6e-5 & 1.9e-2 & 2.2e-2 \\
      JAX-CFD & 7.1e-6 & 1.9e-2 & 2.2e-2 \\
      PhiFlow & 1.0e-5 & 1.9e-2 & 2.2e-2 \\
      PICT & 4.6e-5 & 8.4e-3 & 4.3e-2 \\
      Warp-NS & 3.7e-5 & 4.9e-3 & 5.5e-3 \\
      XLB & 1.1e-5 & 1.9e-4 & 1.8e-3 \\
    \bottomrule
  \end{tabular}
\end{table}

All differentiable solvers reach cosine similarity $>0.999$ against finite
differences, consistent with correct AD. The optimal $\varepsilon$ differs by
orders of magnitude across solvers and domains, so a fixed step size can make
a correct gradient look wrong (\cref{fig:app:fdcheck}).

For the grid-based solvers, the forward discretization limits gradient
accuracy on the Taylor--Green checks. Every solver's error is larger against the
exact gradients than against finite differences, by three orders of magnitude
for the staggered-grid solvers, because a solver's computed gradient and its finite-difference
reference are both impacted by discretization error, while the exact gradient is not.
Optimization with these gradients therefore converges toward the optimum of the discretized problem
rather than the continuous one, and a more accurate gradient would first require a finer discretization.

\begin{figure}[t]
  \centering
  \includegraphics[width=\linewidth]{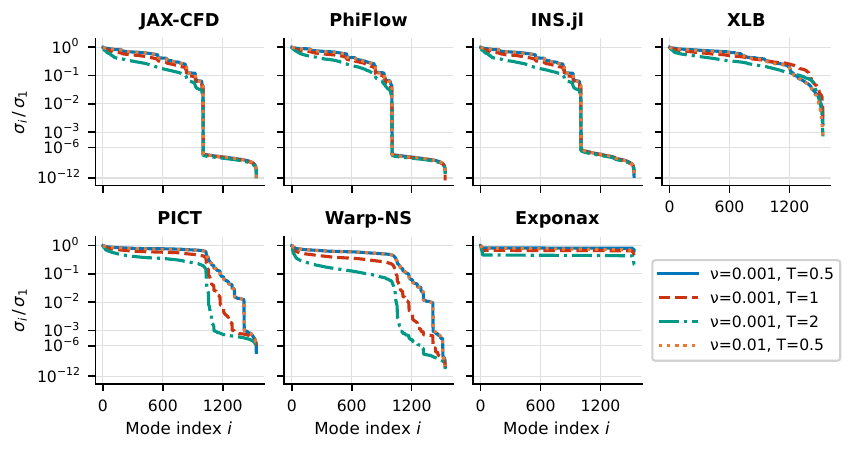}
  \caption{Normalized Jacobian singular value spectra ($\sigma_i / \sigma_1$) on F3 ($N=8$) after $T=0.5$, 1, and 2 time units (10, 20, and 40 steps) at $\nu=10^{-3}$, and after $T=0.5$ at $\nu=10^{-2}$.
    Exponax and XLB keep a nearly flat spectrum at every rollout length. JAX-CFD, INS.jl, and PhiFlow drop to round-off after about two thirds of the modes, and the spectra of PICT and Warp-NS steepen as the rollout grows.}
  \label{fig:conditioning}
\end{figure}

The three shapes of the F3 Jacobian spectra (\cref{fig:conditioning}) mostly
reflect how each solver treats initial perturbations that are not
divergence-free and thus violate incompressibility~\citep{guermond2006overview,bhatia2013helmholtz}. Solvers that project
them out exactly have singular values at round-off, so their condition numbers
reflect round-off rather than the flow (\cref{app:gradient}). The spectra did
not indicate which solvers stay stable over long rollouts, since each shape
occurs both among solvers that stay stable and among those that return
non-finite gradients (\cref{sec:scaling}).

\subsection{Computational cost and scaling}
\label{sec:scaling}

\begin{figure}[t]
  \centering
  \includegraphics[width=0.755\linewidth]{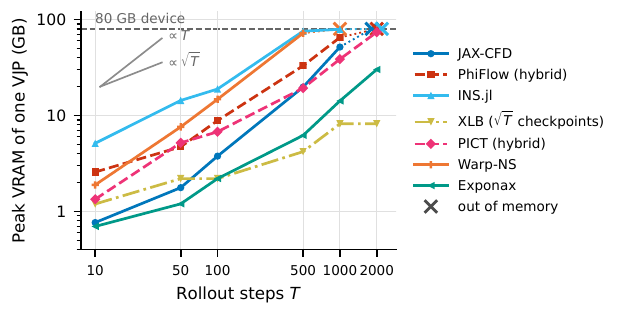}
  \caption{Peak GPU memory of one VJP against rollout steps on F3
    ($N=48$), by differentiation strategy (line style).}
  \label{fig:memory_rollout}
\end{figure}

Absolute solver cost varies by orders of magnitude on the same problem
(\cref{tab:merged}; scaling curves in \cref{app:cost}), even between GPU
solvers that handle the same channel. XLB's explicit lattice Boltzmann
updates need no linear solve, whereas PICT's linear solves are bound by
device synchronization, and its F2 forward solve takes more than 200 times as
long.

Memory use depends on the differentiation strategy. Reverse-mode AD through a
time loop stores one state of size~$S$ for each of its $T$ steps, and hence
needs $O(T\cdot S)$ memory. Checkpointing, used here only by XLB, reduces this
to $O(\sqrt{T}\cdot S)$, and steady adjoints need only $O(S)$.
Unrolling $K$ inner solver iterations multiplies both the stored trajectory
and the cost of the reverse sweep by~$K$. By the implicit function theorem, a
solve-level adjoint instead needs only the converged state and one solve with
the transposed operator, which is why several solvers combine AD with adjoint
solves (AD-S in \cref{tab:solvers}). The F3 measurements follow these
scalings (\cref{fig:memory_rollout}), and without checkpointing, the requirement to store
the entire trajectory limits a $48^3$ rollout to a few thousand steps even with 80\,GB. Memory is not always the
first limit, however. On a longer F3 sweep, PhiFlow, Warp-NS, and XLB return
non-finite gradients after a few hundred steps, long before the other solvers
run out of memory (\cref{fig:horizon_sweep}). The steady mesh problems do not require
storing the trajectory and fail only at millions of elements (\cref{app:cost}).

Reverse-mode AD should cost a small constant multiple of the forward
pass~\cite{griewank2008evaluating}, and most VJPs here cost one to four
forward passes (\cref{tab:app:ratios}). This makes the ratio a cheap
diagnostic for integration problems. During development, we found ratios of up to
$1\,100$, which exposed three integration bugs, all fixed in the reported results
(\cref{app:ratios}). INS.jl's ratio stays high at every grid size, likely because
Zygote cannot differentiate the library's in-place time stepper, so its VJP
runs a slower non-mutating rollout instead.

\subsection{Optimization task performance}
\label{sec:optimization}

\begin{figure}[t]
  \centering
  \includegraphics[width=\linewidth]{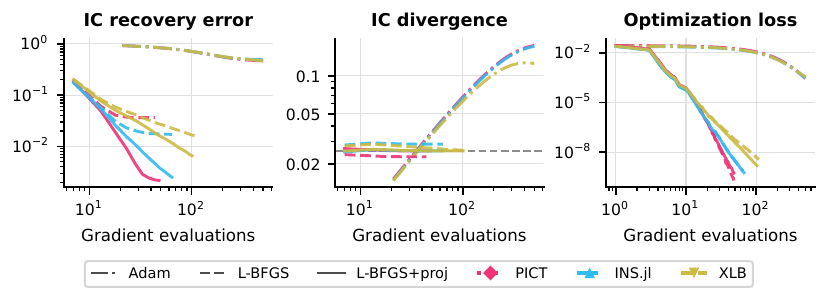}
  \caption{F3 IC recovery for selected solvers and different optimizers. The IC
    error is the relative $\ell_2$ distance between the recovered and true initial
    states. Full version in \cref{fig:app:recovery}.}
  \label{fig:recovery_main}
\end{figure}

Every solver is run with the same hyperparameters and iteration budget per
task, so differences in convergence are attributable to the solver's
gradient alone. On every task, the solvers that can run it reach
comparable results (\cref{app:recovery}). They converge to the same optimum
on S, and drag drops by 46 to 48\% on F2. On H, the misfit drops by about eight orders of magnitude with every solver,
but because the inversion is poorly conditioned,
the recovered conductivity depends on the optimizer's path. Two solvers cannot run the F2 channel:
Exponax represents only periodic fields, and the JAX-CFD API cannot express the channel's inflow,
outflow, and obstacle.

On F3, the choice of optimizer matters more than the choice of solver
(\cref{fig:recovery_main}). L-BFGS~\cite{byrd1995limited} recovers the initial state an order of
magnitude more accurately than Adam~\cite{kingma2015adam}, in a fraction of the gradient
evaluations. Projecting the gradient onto divergence-free fields before each
L-BFGS update (L-BFGS+proj) improves the recovery on every solver, likely because it
keeps components not compatible with incompressible flow out of the optimizer's
curvature model.

\subsection{Solver-in-the-loop training}
\label{sec:sil}

The benchmark checks gradients outside of any training loop. To test
whether these checks predict how a solver behaves during training, we set up a
small periodic CNN to correct each solver's coarse ($64^2$) run of forced
Kolmogorov flow against the solver's own $256^2$ run, in three variants:
supervised one-interval pairs, a 16-interval rollout with the gradient
stopped at every solver call, and the same rollout differentiated through
the solver (SIL). Four solvers run the full study (JAX-CFD, INS.jl, XLB, and
Exponax), and the setup and reasons for exclusions are outlined in \cref{app:sil}.

During the study, we found one gradient defect, PICT's adjoint tolerance, which
shows up in the chained gradient check but not in single-call checks
(\cref{app:kolmogorov}). For the four solvers that run the study, whose
gradient errors are all below $2\times10^{-3}$, the test errors track forward
accuracy rather than gradient accuracy (\cref{tab:app:sil}). The error drops
most for the two solvers with the larger uncorrected error, and by less for the
other two. Because each corrector is trained against its own reference and starts from
a different error, these gains are hard to compare across solvers.

Averaged over three seeds, supervised training reached the lowest test error
on every solver except Exponax, and SIL's average never fell below the
stopped-gradient variant's, although individual SIL seeds did. Earlier work found
training through the solver to outperform supervised
training~\cite{um2020sol} and, by a smaller margin, the stopped-gradient
rollout, also on Kolmogorov flow~\cite{list2024differentiability}. Our
corrector has about 2\,600 parameters and is trained for 256 updates without a
curriculum or per-variant tuning, so this is not a fair test of that advantage.
Within this setup, the solvers differed under SIL training only in cost.

\section{Discussion and conclusion}
\label{sec:conclusion}

Gradient accuracy, computational cost, memory, stability, and setup
compatibility vary widely across solvers. Some of these properties
determine whether a solver can serve a workflow at all, and others only how fast it does so. Based on our
benchmark tasks, we highlight five practitioner-facing observations.

\paragraph{Gradient quality is often good enough}
On every task where they can run, all solvers that produce gradients
reach comparable optima despite order-of-magnitude differences in
gradient accuracy. More complex tasks may make gradient quality limiting,
but these benchmarks suggest that optimization and training are often
robust to it. Solvers failed instead because they could not express the
task, turned unstable over long rollouts, or ran out of memory.

\paragraph{Rollout length is the critical scaling parameter}
For unrolled training and long-horizon control, two limits on rollout
length can make a solver unusable. Without checkpointing, memory limits most
F3 solvers at a few thousand steps on a $48^3$ grid, and three of seven solvers
return non-finite gradients after a few hundred steps (\cref{sec:scaling}). Both aspects should
be tested at the intended rollout length before committing to a solver.

\paragraph{On steady linear problems, pick the fastest (or friendliest) solver}
On problems H and S, all solvers produce equivalent
results regardless of AD strategy or numerical method, so the choice
reduces to forward speed at the intended size and integration effort.

\paragraph{Reverse-mode AD offers the best effort-to-capability tradeoff}
Hand-derived adjoints can be cheap, as TopOpt.jl's self-adjoint compliance
gradient costs about one forward solve (\cref{tab:merged}), but each new
control variable or objective requires new derivations. Reverse-mode AD
through modern frameworks produces gradients accurate enough for
optimization at a fraction of the implementation cost. At the sizes of
\cref{tab:merged}, AD gradients cost 1.4 to 5.3 times the forward time.

\paragraph{Measure rather than predict}
Some properties can be anticipated from how a solver works: memory scales with the differentiation strategy, gradient
accuracy is bounded by the discretization, and Fourier methods are
restricted to periodic domains. Others, including some that determined whether a solver was usable here, could not.
Solvers of the same numerical family and AD strategy ended up on both sides of the stability limit, their Jacobian
spectra did not predict stability (\cref{sec:quality}), and gradient defects
and the largest cost outliers came primarily from how a solver was built or wrapped.
Practitioners should therefore test candidate solvers on their own task, for
example by adapting one of Mosaic's task templates (\cref{sec:scope}).

\paragraph{Limitations}
Our conclusions rest on three choices. The gradient reference is central
finite differences at the best perturbation size, corroborated by the exact
Taylor--Green gradients. Each solver runs as published, with every
departure from its default usage recorded, and expert tuning of
resolution, time step, or preconditioner could narrow the gaps we observe.
Standalone optimization tasks stand in for broader pipelines, which the
solver-in-the-loop study supports for one setup.

The scope is deliberately narrow. The tasks are easy (small grids, moderate
Reynolds numbers, short rollouts, steady linear mesh problems), and only
the solver-in-the-loop study includes learned components. Results should
not be extrapolated to harder problems, such as fully developed 3D
turbulence, long-horizon optimization, or nonlinear mechanics, or to other
hybrid pipelines without further validation. All timings come from a
single A100 machine, and rankings may shift on other hardware, especially for
float64 solvers such as XLB on consumer GPUs with low float64 throughput.

Reported numbers are a snapshot of pinned solver versions, and rankings will
change as solvers improve. Tracking such changes is part of the benchmark's
purpose, and with continuous re-evaluation, the published results stay
current. We therefore deliberately state our conclusions in terms of general
failure modes and solver properties, and expect them to hold beyond individual
ranking changes.

\paragraph{Future work}
Natural extensions include additional domains (multiphysics, compressible
flow), solver backends (Enzyme/LLVM-level AD), and hybrid physics-ML tasks.
The infrastructure is in place, and we invite community contributions.

\section*{Acknowledgments}
We thank everyone who has contributed to Mosaic. In particular, we thank the
developers of the integrated solvers who reviewed, corrected, or improved their
solvers' integrations, and everyone who reported issues against the benchmark.

\PLrefheading
\bibliography{refs}

\clearpage

\appendix
\crefalias{section}{appendix}
\crefalias{subsection}{subappendix}

\section{Software and reproducibility}
\label{app:software}

\subsection{License, access, and reproduction}

Mosaic is released under the Apache~2.0 license (individual solver backends
retain their upstream licenses). The repository, documentation, and
reproduction instructions are available at
\url{https://github.com/pasteurlabs/mosaic}.

The paper corresponds to the tag
\href{https://github.com/pasteurlabs/mosaic/tree/v0.3%2Bpaper-repro}{\texttt{v0.3+paper-repro}}
which contains all code, plotting scripts,
solver wrappers, and a pinned dependency lockfile sufficient for full
reproduction. Two reproduction paths are supported. Figures can be
regenerated from the result archive linked in the repository README without
re-running any solver. Alternatively, the full experiment suite can be
re-run from scratch; this requires Docker and, for GPU solvers, an NVIDIA
GPU with the NVIDIA Container Toolkit. The lockfile pins every transitive
dependency to the exact versions used for the paper, and solver containers
are built on-demand by the harness. Step-by-step instructions for both
paths are in the repository README.

\subsection{Architecture}

Each solver backend is a \emph{Tesseract}~\cite{haefner2025tesseract}: a
Docker container that exposes a typed forward map (\texttt{apply}), shape
inference (\texttt{abstract\_eval}), and optionally a VJP
(\texttt{vector\_jacobian\_product}) over Pydantic schemas. The container
boundary isolates solver-specific dependencies (JAX, PyTorch, Julia, C++,
Warp) from one another and from the evaluation harness.
\texttt{tesseract-jax} bridges the container interface into JAX's
\texttt{jit}/\texttt{grad} transform chain, so the harness can
differentiate through any backend regardless of its native AD framework
using a single code path.

Each backend is defined by three files: a Python module implementing the
entry points above, a YAML configuration declaring solver metadata (name,
AD strategy, backend framework, plot style), and a requirements file listing
dependencies. The evaluation harness discovers solvers automatically from
their configuration metadata and runs the benchmark suites (forward
accuracy, gradient quality, computational cost, and optimization
convergence) without solver-specific code.

\subsection{Extensibility}

Contributing a new solver requires implementing the three files described
above for an existing domain. The harness discovers the backend
automatically from its configuration metadata; no registration step or
evaluation-code changes are needed. A step-by-step tutorial (adding an LBM
solver to the fluid domain) is included in the repository documentation.
New benchmark domains can be scaffolded from built-in templates and require
interface schemas, at least one solver backend, and a problem configuration
defining initial conditions, error metrics, and suite defaults. Existing
solver configurations can also be improved via pull requests; CI re-runs
the full evaluation suite and posts a before/after comparison.

\subsection{Maintenance and versioning}

We maintain the evaluation protocol, the Tesseract interface contract, CI
infrastructure, and the paper-pinned solver set (tagged release, kept
buildable). New solver contributions, domain-specific task formulations,
and upstream solver maintenance are community-owned. If an upstream solver
breaks due to a dependency update, we document version pins and accept
community fixes but do not patch upstream code ourselves.

Each release is tagged with a semantic version. Continuous integration runs
the benchmark suite on every pull request to
\texttt{main} that touches a solver or the harness, committing results
to a dedicated branch with a machine-readable diff against the previous
baseline. For tagged releases, CI runs the full suite and publishes a result
archive as a GitHub release artifact.

\section{Extended experimental results}
\label{app:extended}

\subsection{Benchmark specifications}
\label{app:effort}

\paragraph{Integration effort}
The \textbf{IO}, \textbf{Num.}, and \textbf{Conf.}\ columns in \cref{tab:merged}
report bespoke lines of code by category.
Each wrapper's code is tagged with \texttt{mosaic:<category>} comments, and
\texttt{paper/count\_loc.py} in the repository counts the lines per tag and maps them
to the columns: \texttt{io} to IO, \texttt{init} to Conf., and \texttt{physics},
\texttt{grad}, and \texttt{util} to Num.
\emph{IO} covers serialization and file-based data exchange with external solver processes.
\emph{Num.}\ is problem-specific numerical code (assembly, time-stepping, boundary
conditions, and gradient code), and the gradient lines quoted in \cref{sec:baselines}
are its \texttt{grad} part.
PhiFlow, JAX-CFD, and Exponax each need at most 20 gradient lines for a single
VJP that covers all differentiable inputs. INS.jl (Zygote) needs 228, PICT
(PyTorch) 170, and XLB (JAX) 91, half of them for checkpointing its rollout.
The Warp-NS wrapper contains the entire flow solver, hence its over 2\,400
numerical lines. Of these, 235 are gradient code, the most of any solver, even
though Warp generates the adjoint of each kernel automatically.
\emph{Conf.}\ covers solver initialization and mesh setup.
Glue code (Tesseract schema definitions and entry-point boilerplate) is excluded; it
consistently amounts to 100--400\,LOC across all backends.

\paragraph{Error column definitions}
\textit{Ref.}: $\|\mathbf{f}_{\text{solver}} - \mathbf{f}_{\text{ref}}\|_2 / \|\mathbf{f}_{\text{ref}}\|_2$,
where $\mathbf{f}_{\text{ref}}$ is a checked-in consensus of all solvers on the multimode
flow (F2), a converged Exponax run at $N=64$ spectrally truncated to the benchmark grid
(F3), or a checked-in trimmed mean of all solvers' compliances (S, H).
This is a forward accuracy metric, not a gradient comparison.
\textit{An.}: $\|\mathbf{f}_{\text{solver}} - \mathbf{f}_{\text{TGV}}\|_2 / \|\mathbf{f}_{\text{TGV}}\|_2$,
where $\mathbf{f}_{\text{TGV}}$ is the TGV closed-form forward solution (F2 only).
\textit{FD}: for $K$ random directions $\mathbf{d}_k$, the error
$|D_k - \langle \mathbf{g}, \mathbf{d}_k\rangle| / \max(|D_k|, |\langle \mathbf{g}, \mathbf{d}_k\rangle|)$
of the solver's gradient $\mathbf{g}$ of $J(\mathbf{x})=\sum \mathbf{f}(\mathbf{x})^2$
against the central difference
$D_k = [J(\mathbf{x}+\varepsilon s\mathbf{d}_k) - J(\mathbf{x}-\varepsilon s\mathbf{d}_k)]/(2\varepsilon s)$,
where $s$ is the root mean square of the input $\mathbf{x}$. We report the median over
directions at the best $\varepsilon$ in $\{5, 1, 10^{-1}, \ldots, 10^{-4}\}$ (F2, F3),
$\{2, \ldots, 10^{-4}\}$ (S), or $\{1, \ldots, 10^{-4}\}$ (H).

\paragraph{Hardware}
All experiments ran on an Azure \texttt{Standard\_NC24ads\_A100\_v4} instance
(24 vCPUs of an AMD EPYC 7V13, 216\,GiB RAM, one NVIDIA A100 80\,GB PCIe), and
CPU-only solvers ran on the same instance's CPU.
The default suites took about 13.5 hours on this instance, the extended mesh
sweeps about 22 hours, the optimizer comparison of \cref{app:recovery} about
10 hours, and the solver-in-the-loop study about 3 hours.
The results on the project website are regenerated by continuous integration
on NVIDIA T4 runners, so their absolute times differ from the ones reported here.

\paragraph{Timing and memory measurement}
Each measurement starts with one untimed warmup call, during which JIT
compilation and CUDA kernel caching take place. Timed trials follow, 3 on F2 and F3 and 10 on S and H,
because the fastest S and H points take only tens of milliseconds and vary most
between trials. The reported value is
the median, which is robust to a single slow trial, and the scaling figures show the
fastest and slowest trial as error bars.
A forward time covers one \texttt{apply} request and a VJP time one
\texttt{vector\_jacobian\_product} request, whose cotangent comes from an untimed
\texttt{apply} beforehand. Every solver recomputes its forward solution inside the
VJP request, so the VJP time is the cost of one gradient from scratch: a forward
plus an adjoint pass. TopOpt.jl's compliance gradient is self-adjoint and needs no
second solve, so its VJP costs about one forward solve.
The timed window covers host-side serialization, the Tesseract HTTP round-trip, solver
compute, and response deserialization; container startup is excluded.
If a solver's first trial at some size takes longer than 1\,000\,s, its sweep stops there,
and every request to a solver container times out after 7\,200\,s.
Peak VRAM is NVML's used memory of the whole device over the warmup and the trials.
It includes the CUDA context and the allocator pools and excludes the 768\,MiB the
driver reserves. Peak RAM is the solver container's cgroup memory as reported by
Docker. Allocator pools persist while a container serves a sweep, which runs in
increasing size, so each recorded peak is the running maximum of the true per-size
peaks. All JAX-based solvers run with the same allocator settings (no preallocation,
memory fraction 0.95).

Each domain's experiments use the following settings.

\paragraph{F2 (2D NS, grid)}
Periodic domain $[0,2\pi)^2$ on a uniform Cartesian grid, except for the
cylinder experiments, which use the Sch\"afer--Turek DFG 2D-1
channel~\cite{schafer1996benchmark}: $[0,2.2]\times[0,0.41]$ with a cylinder of
diameter $D=0.1$ at $(0.2,0.2)$, a parabolic inflow of peak speed $U_m=0.3$
(mean $\bar{U}=0.2$), a zero-gradient outflow at $p=0$, no-slip walls, and
$\nu=10^{-3}$, so that $\mathrm{Re}=\bar{U}D/\nu=20$.
\begin{itemize}\itemsep0pt
  \item \textbf{Fwd/VJP}: $64\times64$ grid (4\,096 cells), multimode IC,
        $\nu=0.01$, $\Delta t=0.01$, 100 steps.
  \item \textbf{Cost}: $N\in\{64,128,192,256\}$ (4\,k--66\,k cells) at 100 steps.
  \item \textbf{Error (Ref., An.)}: $64\times64$ grid, $\nu=0.01$,
        $\Delta t=0.05$, 20 steps. Ref.\ uses the multimode IC, and An.\ the TGV
        against the closed form $\mathbf{u}(t)=\mathbf{u}(0)e^{-2\nu t}$.
  \item \textbf{Error (FD)}: $16\times16$ grid, multimode IC, $\nu=0.001$,
        $\Delta t=0.05$, 20 steps, 20 random directions.
  \item \textbf{Convergence}: drag minimization in the DFG channel at 10 cells
        per diameter to $t=4$. The control is the inlet profile (41 cells),
        projected onto non-negative profiles with the benchmark's flow rate and
        a peak of at most $U_m$. The loss is $C_d/C_{d,0}$ relative to the
        parabolic profile.
\end{itemize}

\paragraph{F3 (3D NS, grid)}
Periodic domain $[0,2\pi)^3$ on a uniform Cartesian grid.
\begin{itemize}\itemsep0pt
  \item \textbf{Fwd/VJP}: $32^3$ grid (32\,768 cells), TGV IC, $\nu=0.01$,
        $\Delta t=0.01$, 50 steps.
  \item \textbf{Cost}: $N\in\{16,32,48,64,96,128\}$ (4\,k--2.1\,M cells) at 50 steps.
  \item \textbf{Error (Ref.)}: $16^3$ grid, TGV IC, $\nu=0.01$, $\Delta t=0.01$,
        50 steps, against a converged Exponax run at $N=64$ truncated to $N=16$
        (truncation changes it by less than $10^{-5}$).
  \item \textbf{Error (FD)}: $16^3$ grid, TGV IC, $\nu=0.001$, $\Delta t=0.05$,
        10 steps, 10 random directions.
  \item \textbf{Convergence}: recovery of a random divergence-free IC
        (\cref{app:ics}) on a $16^3$ grid from its state after 100 steps
        ($\nu=0.01$, $\Delta t=0.02$), starting from zero, for three target
        fields (seeds 0--2).
\end{itemize}
The reference is a converged run rather than the closed form because the 3D TGV
is not advection-free, so its closed-form decay holds only over the first few
steps. A converged pseudo-spectral solution differs from it by
$2.5\times10^{-3}$ after one step and by 0.12 after 50 steps.

\paragraph{S (Structural mechanics, mesh)}
Cantilever beam $[0,2]\times[0,1]\times[0,1]$ with HEX8 elements and SIMP
penalization ($p=3$, $E_\text{max}=70\,000$\,MPa, $\nu_\text{mat}=0.3$,
$x_\text{min}=10^{-3}$).
\begin{itemize}\itemsep0pt
  \item \textbf{Fwd/VJP}: $128\times2\times64$ mesh (${\approx}16\text{k}$
        elements), uniform density $\rho_0=0.5$, distributed unit load $F=1$.
  \item \textbf{Cost}: $n_x\times2\times n_x/2$ meshes with
        $n_x\in\{4,\ldots,128\}$ (16--16\,k elements).
  \item \textbf{Error (FD)}: $8\times2\times4$ mesh (64 elements), random
        density, corner point load, 6 random directions.
  \item \textbf{Convergence}: compliance minimization, $16\times2\times8$ mesh
        (256 elements), volume fraction $0.5$.
\end{itemize}

\paragraph{H (Heat transfer, mesh)}
Quasi-2D heated slab $[0,2]\times[0,1]$ with a single layer of HEX8 elements
and SIMP penalization ($p=3$, $k_\text{max}=1$, $k_\text{min}/k_\text{max}=10^{-3}$).
\begin{itemize}\itemsep0pt
  \item \textbf{Fwd/VJP}: $256\times128\times1$ mesh (${\approx}33\text{k}$
        elements), uniform SIMP density $\rho_0=0.5$, Neumann flux
        $Q_\text{total}=1$ uniformly over the right face, and $T=0$ at $x=0$.
  \item \textbf{Cost}: $n_x\times n_x/2\times1$ meshes with
        $n_x\in\{16,\ldots,1024\}$ (128--524\,k elements).
  \item \textbf{Error (FD)}: $8\times4\times1$ mesh (32 elements), random
        conductivity, 6 random directions.
  \item \textbf{Convergence}: conductivity recovery of a two-Gaussian target,
        $16\times8\times1$ mesh.
\end{itemize}

\subsection{Initial conditions}
\label{app:ics}

Each benchmark domain uses a fixed set of initial conditions, shared across
all solvers.

\begin{figure}
  \centering
  \includegraphics[width=\linewidth]{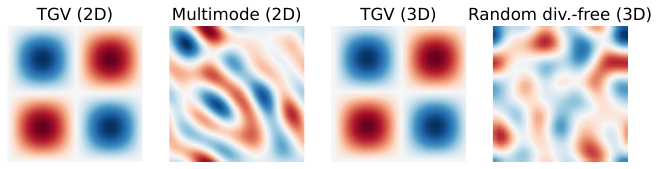}
  \caption{Initial conditions for the F2 and F3 domains (vorticity; 3D shown as $z{=}0$ slice).}
  \label{fig:app:ics_ns}
\end{figure}

\paragraph{Incompressible NS (\cref{fig:app:ics_ns})}
\textbf{Taylor--Green vortex (TGV, 2D):}
$u = \sin(x)\cos(y)$, $v = -\cos(x)\sin(y)$ on $[0,2\pi)^2$.
Admits the analytic solution $\mathbf{u}(t)=\mathbf{u}(0)e^{-2\nu t}$,
enabling exact error measurement without a reference solver.
\textbf{Multimode:} solenoidal field constructed from a spectral-space
stream function with energy concentrated in a ring at wavenumber $k=2$
($\sigma_k=0.5$) and random phases; peak speed normalized to~0.3.
\textbf{DFG inflow:} the parabolic inlet profile $u_x(y)=4U_m y(H-y)/H^2$ of the
DFG 2D-1 channel ($U_m=0.3$, $H=0.41$), the inflow of the cylinder experiments
and the starting point of the drag optimization.
\textbf{Taylor--Green vortex (TGV, 3D):}
$u = \sin(x)\cos(y)\cos(z)$, $v = -\cos(x)\sin(y)\cos(z)$, $w = 0$
on $[0,2\pi)^3$.
Evolves into a vortex-dominated turbulent state under the NS dynamics.
\textbf{Random divergence-free (3D):} the curl of a random spectral vector
potential with energy in a shell at $|k|=2$ (width~1), normalized to unit peak
speed. It is the target of the F3 recovery task, with one field per seed.

\paragraph{Structural mechanics}
Both ICs are material density fields over the beam mesh.
\textbf{Uniform:} $\rho_0 = 0.5$ everywhere.
\textbf{Random:} Gaussian noise centered at $\rho_0=0.5$ ($\sigma=0.3$),
clipped to $[0.05,\,0.95]$, the starting point of the gradient check.

\paragraph{Heat transfer}
All ICs are conductivity fields over the thermal mesh.
\textbf{Uniform conductivity:} SIMP density $\rho_0 = 0.5$ everywhere.
\textbf{Random conductivity:} Gaussian noise in the SIMP density centered at $\rho_0=0.5$
($\sigma=0.3$), clipped to $[0.05,\,0.95]$, the starting point of the gradient check.
\textbf{Two-Gaussian conductivity:} ground-truth field for the conductivity
recovery task; two Gaussian peaks at $(0.3\,L_x,\,0.5\,L_y)$ and
$(0.7\,L_x,\,0.5\,L_y)$ with width $\sigma=0.15\,\min(L_x,L_y)$.

\subsection{Cost}
\label{app:cost}

\begin{figure}
  \centering
  \includegraphics[width=\linewidth]{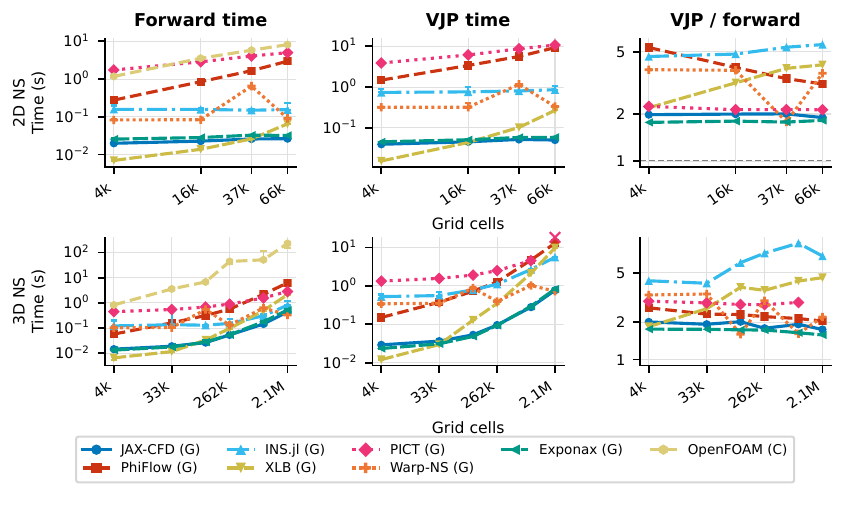}
  \caption{Gradient overhead scaling for the fluid domains (2D NS, top;
    3D NS, bottom). Left: forward time; center: VJP time; right:
    VJP/forward ratio (dashed line marks 1:1). Points are medians over the
    timed trials, and error bars span the fastest and slowest trial. All axes are
    log-log. (G)\,/\,(C) denotes GPU\,/\,CPU execution.}
  \label{fig:app:scaling}
\end{figure}

\begin{figure}
  \centering
  \includegraphics[width=\linewidth]{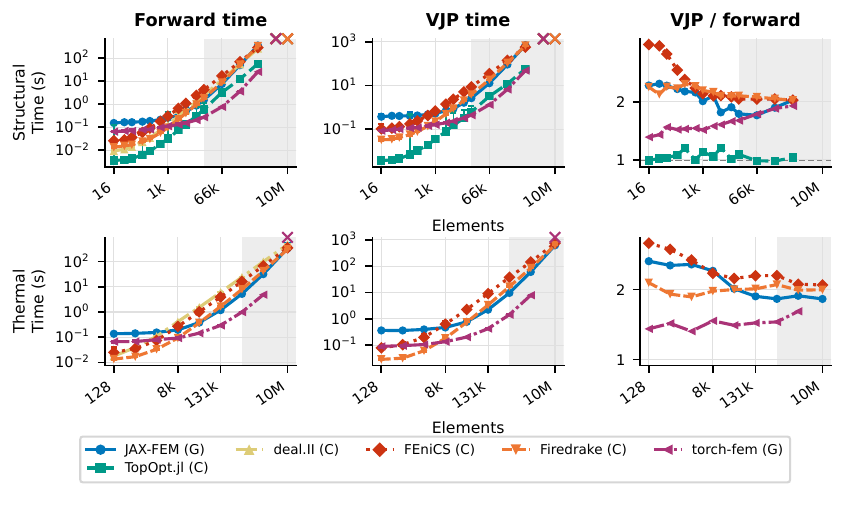}
  \caption{Gradient overhead scaling for the mesh domains (Structural,
    top; Thermal, bottom). Same layout as \cref{fig:app:scaling}.
    deal.II VJP times are omitted (no native adjoint). Shaded: sizes beyond
    the main sweep, from the extended sweeps with three trials per point.
    A cross at the top edge marks a failed size; \cref{app:cost} gives the
    causes.}
  \label{fig:app:scaling:fem}
\end{figure}

\Cref{fig:app:scaling,fig:app:scaling:fem} show forward time, VJP time, and VJP/forward ratio vs.\ element count for all four domains.
The x-axis is total element count, parameterized by a single integer $N$:
F2 uses an $N\times N$ uniform Cartesian grid ($N\in\{64,128,192,256\}$, giving $4\text{k}$--$66\text{k}$ cells);
F3 uses an $N\times N\times N$ Cartesian grid ($N\in\{16,32,48,64,96,128\}$, giving $4\text{k}$--$2.1\text{M}$ cells);
S uses an $N\times 2\times\lfloor N/2\rfloor$ hexahedral mesh ($N\in\{4,\ldots,128\}$, giving $16$--$16\text{k}$ elements);
H uses an $N\times\lfloor N/2\rfloor\times 1$ hex mesh ($N\in\{16,\ldots,1024\}$, giving $128$--$524\text{k}$ elements).
For the mesh domains, \cref{fig:app:scaling:fem} adds extended sweeps from a separate run on the same machine, which continue S to $N=3\,200$ and H to $N=4\,500$ (10\,M elements each) with three trials per point.
Table~\ref{tab:merged} wall times are at $N=64$ (F2), $N=32$ (F3), $N=128$ (S), $N=256$ (H).
The rollout-length sweeps run F2 at $N=128$ for 10--1\,000 steps and F3 at $N=48$ for 10--2\,000 steps.

The fixed RPC overhead (host serialization and HTTP round-trip) adds approximately 2--20\,ms per call and is visible as a floor for the fastest solvers at small problem sizes.
Lines that terminate before the full size range indicate that the solver either raised an exception (typically an out-of-memory error) or that its first trial exceeded the 1\,000\,s per-trial wall limit at that problem size. All larger sizes for that solver are omitted.
PICT's F3 VJP runs out of memory at $N=128$ after two of three trials, its container exiting with the GPU's memory full and no error message.

On the extended mesh sweeps, all H configurations reach $N=4\,500$ except torch-fem, which runs out of GPU memory at that size.
On S, every solver reaches $N=1\,024$ and none reaches 2\,048, where JAX-FEM and torch-fem run out of GPU memory and the CPU solvers reach the 195\,GB cap of their containers.
At $N=3\,200$, Firedrake fails below the cap because the stiffness matrix has $2.9\times10^9$ nonzeros, more than PETSc's 32-bit indices allow, and FEniCS fails in PETSc's matrix preallocation, probably for the same reason.
Where the extended sweeps overlap the main sweeps, their mean times agree within 10\% except at three points, which differ by up to 22\% because of single slow trials.

\begin{figure}
  \centering
  \includegraphics[width=\linewidth]{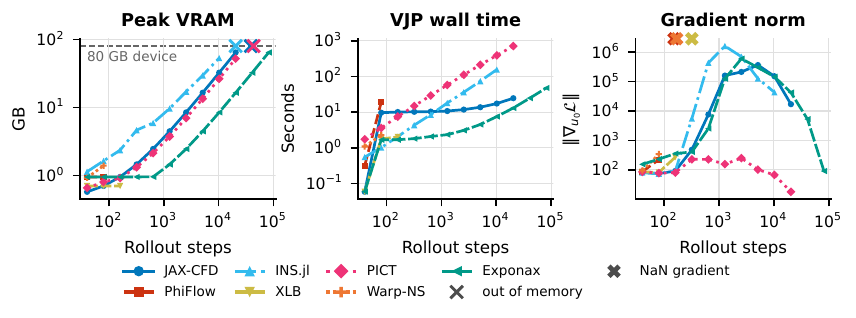}
  \caption{VJP against rollout length on F3 ($N=20$, $\nu=10^{-3}$, $\Delta t=0.05$): peak GPU memory, wall time, and gradient norm.
    Each line ends at the solver's last successful horizon. Out-of-memory failures are marked at the device limit, and non-finite gradients at the top of the gradient-norm panel.}
  \label{fig:horizon_sweep}
\end{figure}

In \cref{fig:horizon_sweep}, we extend the F3 rollout to 81\,920 steps at $N=20$, with one VJP per horizon.
PhiFlow and Warp-NS return non-finite gradients after 160 steps and XLB after 320, while INS.jl, JAX-CFD, and PICT stay stable until they run out of memory after 20\,000 to 41\,000 steps, and Exponax completes all 81\,920.
Two memory readings are less informative than the others. INS.jl's peak is dominated by the CUDA.jl memory pool, and Warp-NS's peak does not track the horizon, because its memory pools carry over from one horizon to the next.

\subsection{VJP/forward ratios}
\label{app:ratios}

\Cref{tab:app:ratios} lists the ratio of VJP to forward time at the \cref{tab:merged} sizes and its range over each sweep.
Early measurements gave ratios of up to 1\,100, which we traced to three integration problems, all fixed in the reported results.
The JAX solvers had fallen back to the CPU, the VJP timing included a second forward solve, and XLB's checkpointing stored the full trajectory.
XLB stays within the expected range, at 2.2 (F2) and 2.6 (F3) at the \cref{tab:merged} sizes and at most 4.6 across grid sizes. Only on the longest F3 rollouts does it reach 7.0.
Two other solvers exceed $5\times$.
INS.jl reaches 6.0--8.6 on the F3 grid sweep at $N\geq48$ and up to 6.3 on F2. Its forward call runs the library's in-place time stepper, but Zygote cannot differentiate array mutation, so its VJP runs a hand-written non-mutating rollout instead. On the CPU, that rollout alone took 2.49\,s against 0.56\,s for the in-place stepper at $N=256$ and 100 steps. The ratio is therefore taken between two different forward implementations.
PhiFlow's 5.3 at F2 $N=64$ falls to 3.1 at $N=256$, consistent with a fixed per-call overhead.

\begin{table}[t]
  \centering\small
  \caption{Ratio of VJP to forward wall time per solver, from median times: at the
    \cref{tab:merged} size, and the range over the grid-size sweep and, for the fluid
    domains, the rollout-length sweep. The VJP request includes a forward pass, so about~2 is
    the typical value for a solver whose gradient costs one adjoint solve.}
  \label{tab:app:ratios}
  \begin{tabular}{@{}l c c c@{}}
    \toprule
    \textbf{Solver} & \textbf{Table size} & \textbf{Grid sweep} & \textbf{Rollout sweep} \\
    \midrule
    \multicolumn{4}{@{}l}{\textit{F2: table size $N=64$; rollout sweep at $N=128$}} \\
      Exponax & 1.8 & 1.8--1.8 & 1.5--1.9 \\
      INS.jl & 4.7 & 4.7--5.6 & 4.1--6.3 \\
      JAX-CFD & 2.0 & 1.9--2.0 & 1.3--2.3 \\
      PhiFlow & 5.3 & 3.1--5.3 & 3.4--5.2 \\
      PICT & 2.2 & 2.1--2.2 & 2.1--2.3 \\
      Warp-NS & 3.8 & 1.8--3.8 & 2.9--4.5 \\
      XLB & 2.2 & 2.2--4.1 & 1.6--4.3 \\
    \midrule
    \multicolumn{4}{@{}l}{\textit{F3: table size $N=32$; rollout sweep at $N=48$}} \\
      Exponax & 1.8 & 1.6--1.8 & 1.5--2.1 \\
      INS.jl & 4.1 & 4.1--8.6 & 3.8--7.1 \\
      JAX-CFD & 1.9 & 1.7--2.0 & 1.6--2.5 \\
      PhiFlow & 2.3 & 2.0--2.6 & 2.1--2.7 \\
      PICT & 2.9 & 2.8--3.0 & 2.6--2.8 \\
      Warp-NS & 3.4 & 1.6--3.4 & 1.5--1.7 \\
      XLB & 2.6 & 1.8--4.6 & 2.3--7.0 \\
    \midrule
    \multicolumn{4}{@{}l}{\textit{S: table size $n_x=128$}} \\
      FEniCS & 2.1 & 2.1--4.0 & $-$ \\
      Firedrake & 2.1 & 2.1--2.5 & $-$ \\
      JAX-FEM & 1.7 & 1.7--2.5 & $-$ \\
      TopOpt.jl & 1.1 & 0.99--1.2 & $-$ \\
      torch-fem & 1.6 & 1.3--1.6 & $-$ \\
    \midrule
    \multicolumn{4}{@{}l}{\textit{H: table size $n_x=256$}} \\
      FEniCS & 2.2 & 2.2--3.2 & $-$ \\
      Firedrake & 2.0 & 1.9--2.1 & $-$ \\
      JAX-FEM & 2.0 & 1.8--2.6 & $-$ \\
      torch-fem & 1.4 & 1.3--1.5 & $-$ \\
    \bottomrule
  \end{tabular}
\end{table}

\subsection{Timing variance}
\label{app:variance}

\Cref{tab:app:variance} lists the coefficient of variation (CV) of the timed trials per configuration, with the sample standard deviation.
The NS suites use 3 trials per configuration and the mesh suites 10.
Across 513 configurations, the median CV is 2.4\% and 40 configurations exceed 15\%.
Of these 40, 26 belong to the two Julia solvers, TopOpt.jl (12) and INS.jl (14), all at sub-second times. Another 3 are OpenFOAM's F3 runs at $N\geq64$.
Dropping the single slowest trial brings 36 of the 40 below 15\%.

\begin{table}[t]
  \centering\footnotesize
  \caption{Run-to-run variation of the timed cost measurements: coefficient of variation (CV,
    sample standard deviation with $n-1$ over the mean of the timed trials) per configuration,
    i.e.\ per solver, cost experiment, and sweep value, with the 3 configurations of
    highest CV per problem.}
  \label{tab:app:variance}
  \begin{tabular}{@{}l r r r >{\raggedright\arraybackslash}p{0.48\linewidth}@{}}
    \toprule
    \textbf{Problem} & \textbf{Configs} & \textbf{Median CV} & \textbf{CV\,>\,15\%} & \textbf{Highest CV} \\
    \midrule
      F2 & 135 & 1.6\% & 7 & INS.jl fwd 100 steps (65\%); INS.jl fwd 50 steps (41\%); INS.jl fwd $N=256$ (27\%) \\
      F3 & 174 & 1.5\% & 12 & INS.jl fwd $N=32$ (99\%); INS.jl fwd 50 steps (73\%); INS.jl fwd 10 steps (62\%) \\
      S & 132 & 3.9\% & 20 & TopOpt.jl VJP $n_x=12$ (275\%); TopOpt.jl fwd $n_x=4$ (244\%); TopOpt.jl fwd 1 step (212\%) \\
      H & 72 & 3.4\% & 1 & FEniCS fwd $n_x=16$ (17\%); FEniCS VJP 1 step (11\%); FEniCS VJP $n_x=64$ (10\%) \\
    \midrule
      All & 513 & 2.4\% & 40 & \\
    \bottomrule
  \end{tabular}
\end{table}

\subsection{Physical accuracy}
\label{app:physical_accuracy}

The fluid solvers reproduce the expected physics. On the decaying Taylor--Green
vortex, the grid solvers converge at the expected order, apart from one solver
with a known error floor, and the six solvers that run the DFG cylinder benchmark
predict its drag within 3\% of the reference value.

\paragraph{Convergence and agreement}
In a baseline experiment, we refine space and time together at a CFL number of about
0.2 up to $t=1.28$. On F2, the reference is the exact TGV solution
($N\in\{16,32,64,128\}$). There, JAX-CFD, PhiFlow, INS.jl, and OpenFOAM converge
at second order (observed orders 2.0--2.3), from errors of
$3.6$--$4.9\times10^{-2}$ at $N=16$ to below $6\times10^{-4}$ at $N=128$, and
Warp-NS and XLB converge at order 2.0--2.1 from lower errors. PICT levels off
near $2.5\times10^{-3}$, consistent with the splitting error of PISO with two
pressure correctors. Exponax represents the single-mode TGV exactly, leaving
only float32 round-off ($\leq 3\times10^{-5}$). On F3, the reference is a
converged Exponax run ($N\in\{8,16,32\}$), and the grid solvers converge at
observed orders 1.6--2.4. In an agreement experiment, we sweep $\nu$ at fixed
resolution ($N=64$ on F2, $N=16$ on F3). On F3, errors at $N=16$ range from
$3\times10^{-3}$ (Warp-NS) to $4.2\times10^{-2}$ (INS.jl and JAX-CFD), while
Exponax's $5$--$9\times10^{-4}$ is largely its distance from its own finer run.

\begin{figure}
  \centering
  \includegraphics[width=\linewidth]{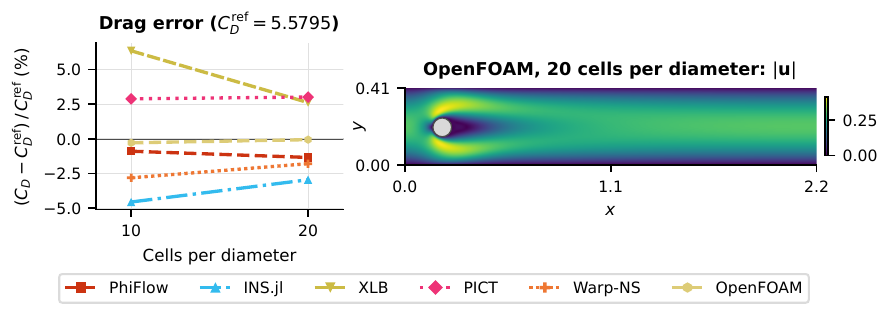}
  \caption{F2 cylinder benchmark (DFG 2D-1, $\mathrm{Re}=20$). Left: signed error of
    each solver's drag coefficient against the reference value $C_d=5.5795$~\cite{nabh1998high} at 10 and 20
    cells per diameter. Right: steady velocity magnitude of the most accurate solver at
    20 cells per diameter.}
  \label{fig:app:cylinder}
\end{figure}

\paragraph{F2 cylinder-flow forward accuracy}
The cylinder experiments use the DFG 2D-1 channel (\cref{app:effort}). In the original
benchmark, Sch\"afer et al.~\cite{schafer1996benchmark} bound the steady drag coefficient to $C_d\in[5.57,5.59]$,
and we score against the high-accuracy reference value $C_d=5.5795$~\cite{nabh1998high}.
Each solver runs at 10 and 20 cells per diameter until $C_d$ is steady, and its $C_d$ at
20 cells per diameter must lie within 5\% of the literature value, a tolerance that
admits staircase representations of the cylinder as well as body-fitted meshes.
Six solvers run it: OpenFOAM and PICT on body-fitted O-grids, PhiFlow, Warp-NS, and XLB
on a shared staircase mask, and INS.jl with Brinkman penalization on the same mask.
At 20 cells per diameter, their $C_d$ errors are 0.05\% (OpenFOAM), 1.3\% (PhiFlow),
1.8\% (Warp-NS), 2.6\% (XLB), 2.9\% (INS.jl), and 3.0\% (PICT; \cref{fig:app:cylinder}).
PICT has no pressure outlet and uses an advective outflow instead, a setup
deviation whose share of its error has not been measured.
Exponax and JAX-CFD cannot represent the channel and do not run it (\cref{app:exclusions}).

\begin{figure}
  \centering
  \includegraphics[width=\linewidth]{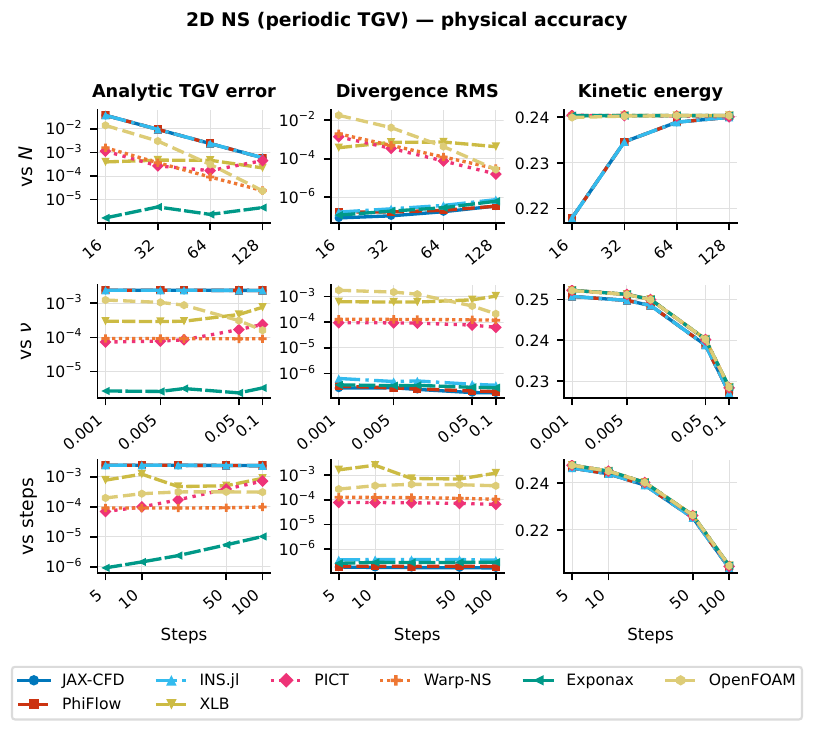}
  \caption{2D NS physical accuracy. Each panel shows one physical metric (y-axis) against one swept parameter (resolution, $\nu$, or steps; x-axis).}
  \label{fig:app:pa_ns2d}
\end{figure}

\paragraph{Physical accuracy sweeps}
The solvers differ most in how strictly they enforce incompressibility.
\Cref{fig:app:pa_ns2d} shows three physical metrics on the 2D TGV over sweeps of resolution $N\in\{16,32,64,128\}$, viscosity $\nu\in\{0.001,0.005,0.01,0.05,0.1\}$, and rollout length (5 to 100 steps).
\textbf{Analytic TGV error} is the relative $\ell_2$ error against the closed-form solution.
\textbf{Divergence RMS} measures how strongly the solver violates the incompressibility constraint $\nabla \cdot \mathbf{u} = 0$, computed with central differences on the output velocity.
\textbf{Kinetic energy} is the domain-averaged $\frac{1}{2}\|\mathbf{u}\|^2$, with the exact $\mathrm{KE}(t) = \tfrac{1}{4}e^{-4\nu t}$ shown as a dashed line.
JAX-CFD, PhiFlow, INS.jl, and Exponax keep the divergence RMS at float32 round-off (below $10^{-6}$), while PICT and Warp-NS lie between $1.6\times10^{-5}$ and $2\times10^{-3}$, falling at second order with resolution.
XLB stays between $4\times10^{-4}$ and $3\times10^{-3}$ without a trend in $N$, consistent with the weak compressibility of the lattice Boltzmann method at the target Mach number of its sub-stepping (\cref{app:included}).

\subsection{Gradient quality}
\label{app:gradient}

\begin{figure}
  \centering
  \includegraphics[width=\linewidth]{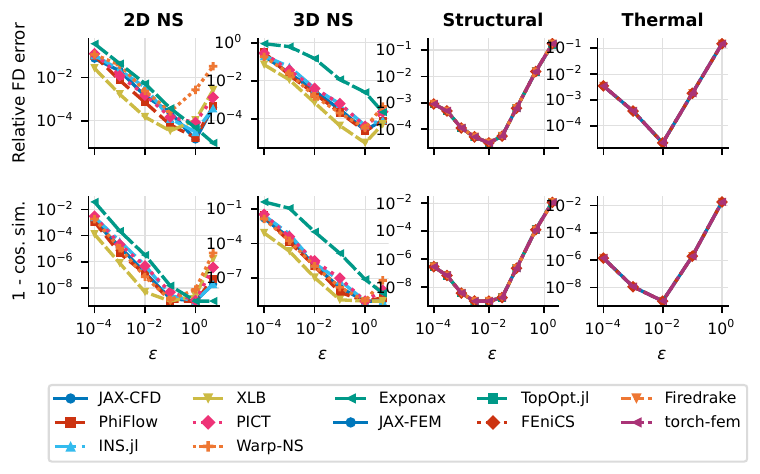}
  \caption{FD gradient verification across all four domains.
    Top row: relative $\ell_2$ error vs.\ perturbation size~$\varepsilon$.
    Bottom row: $1-{}$subspace cosine similarity vs.~$\varepsilon$.}
  \label{fig:app:fdcheck}
\end{figure}

\paragraph{Finite-difference verification}
\Cref{fig:app:fdcheck} shows the FD verification of \cref{sec:quality} for all four domains.
The optimal~$\varepsilon$ shifts noticeably between solver families and between domains: it is larger for spectral solvers (Exponax) than for projection-based solvers, and differs between the structural/thermal and the fluid solvers.
All differentiable solvers reach cosine similarity $>0.999$ at their optimal~$\varepsilon$, consistent with correct gradients in every domain.

\paragraph{Exact-gradient checks}
On F2, we run two checks against the exact Taylor--Green solution at $N=32$ after 20 steps of $\Delta t=0.05$, for $\nu\in\{0.001,0.005,0.01,0.02,0.05\}$.
In the first, we compare the gradient of $J=\sum|\mathbf{u}(T)|^2$ with respect to the initial field against $\nabla J = 2e^{-4\nu T}\mathbf{u}_0$, after projecting out the gradient's curl-free part, since the closed form holds for divergence-free perturbations.
In the second, we compare $\mathrm{d}J/\mathrm{d}\nu$ against $-4TJ$.
Neither requires a step size, but both include the forward discretization error.
Exponax represents the Taylor--Green mode exactly, and its errors of about $10^{-5}$ are consistent with float32 round-off.
JAX-CFD, PhiFlow, and INS.jl average the collocated input onto their staggered grids and back, which scales the mode by $\cos^2(\Delta x/2)$, a factor that enters the gradient twice. At $N=32$ this accounts for their shared initial-state error of $1.9\times10^{-2}$, and their viscosity errors also coincide, at $2.2\times10^{-2}$.
XLB's initial-state error grows with~$\nu$ from $7\times10^{-5}$ to $8\times10^{-4}$, Warp-NS's stays near $5\times10^{-3}$, and PICT's grows from $4\times10^{-3}$ to $2\times10^{-2}$. PICT's viscosity error reaches $5.4\times10^{-2}$ at $\nu=0.001$.
All solvers pass the checks' thresholds (error at most 0.1 and within $3\times$ the peer median). We set these thresholds from preliminary runs on a subset of the solvers.
\Cref{tab:gradchecks} reports the errors at $\nu=0.01$.

\paragraph{Jacobian singular value spectra}
On F3 (\cref{fig:conditioning}), JAX-CFD, INS.jl, and PhiFlow have 1\,002 of 1\,536 singular values above round-off in every configuration, with the rest between about $10^{-8}$ and $3\times10^{-7}\sigma_1$. Their condition numbers of $5\times10^{10}$ to $4.5\times10^{12}$ are therefore set by round-off and vary without a trend in rollout length or viscosity.
Between $T=0.5$ and $T=2$, the number of singular values above $10^{-2}\sigma_1$ falls from 1\,412 to 1\,056 for PICT and from 1\,388 to 1\,044 for Warp-NS. XLB keeps at least 1\,492 of them in every configuration, and Exponax all 1\,536.

\begin{figure}
  \centering
  \includegraphics[width=\linewidth]{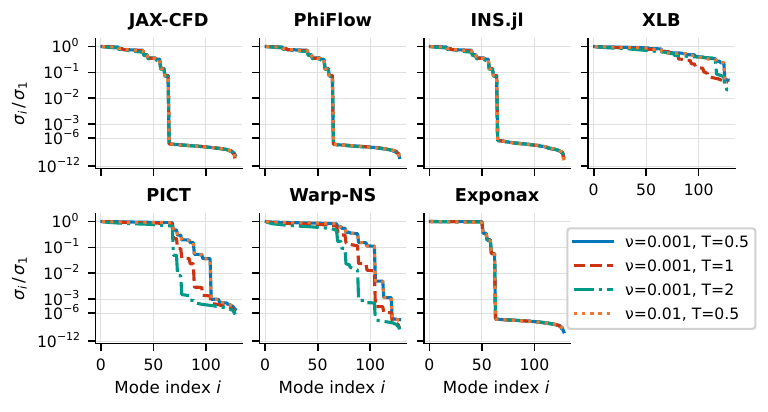}
  \caption{Normalized Jacobian singular value spectra for 2D NS.
    Each panel shows a different $(\nu, T)$ configuration.
    Faster decay corresponds to a more ill-conditioned Jacobian.}
  \label{fig:app:svd2d}
\end{figure}

\Cref{fig:app:svd2d} shows normalized singular value spectra $\sigma_i/\sigma_1$ of the solver Jacobian for 2D NS across varying viscosity and rollout length.
Faster spectral decay indicates a more ill-conditioned Jacobian.
In 2D, unlike on F3 (\cref{fig:conditioning}), rollout length and viscosity change individual condition numbers by up to two orders of magnitude without a consistent direction, while the ordering of the solvers stays the same in every configuration: XLB has the smallest condition number (20--48), followed by PICT ($2\times10^5$--$2\times10^6$), with the other solvers between $2\times10^7$ and $10^{11}$.
Exponax solves the 2D problem in vorticity form and thus discards the curl-free part of its input, so in 2D, unlike in 3D, its condition number is about $10^{10}$, like those of the projection solvers.

\begin{figure}
  \centering
  \includegraphics[width=\linewidth]{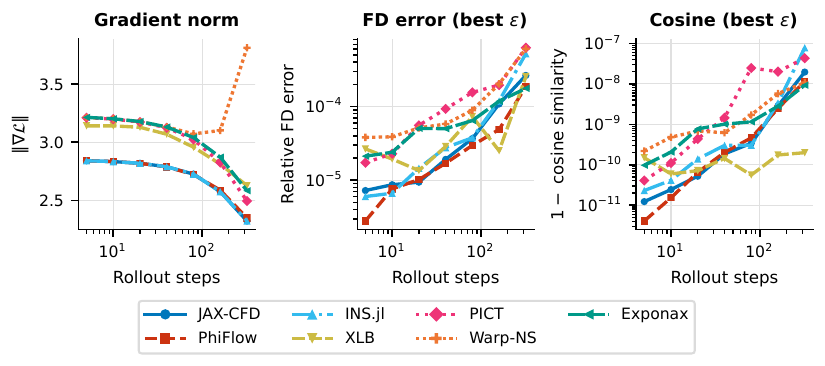}
  \caption{Gradient quality vs.\ rollout length (2D NS).
    Left: gradient norm over rollout length.
    Middle: best-achievable FD relative error at each horizon.
    Right: $1-{}$cosine similarity at the best~$\varepsilon$.}
  \label{fig:app:horizon}
\end{figure}

\paragraph{Gradient quality over rollout length}
\Cref{fig:app:horizon} shows how FD gradient quality evolves with rollout length on the 2D NS domain ($N=16$, 5 to 320 steps).
The best relative FD error grows from $3\times10^{-6}$--$4\times10^{-5}$ at 5 steps to $2\times10^{-4}$--$6\times10^{-4}$ at 320 steps, and $1-{}$cosine stays below $10^{-7}$ throughout, so no solver loses gradient accuracy within this range.
The growth does not follow the conditioning, since XLB, the best-conditioned solver, ends within the range of the others.

\subsection{Optimization}
\label{app:recovery}

\paragraph{Optimizer configurations}
All optimization runs use Adam (\texttt{optax.adam}, default
$\beta_1{=}0.9$, $\beta_2{=}0.999$, $\epsilon{=}10^{-8}$), L-BFGS
(SciPy's L-BFGS-B with its default memory of 10 correction pairs), or, on S,
MMA. Adam runs use patience-based early stopping on the loss:
the run terminates as soon as the loss has not strictly improved over
the running minimum for \emph{patience} consecutive iterations, or when
the maximum iteration budget is reached. L-BFGS stops at the iteration
budget, when the gradient's 2-norm falls below $10^{-7}$, or when its line
search fails.
The x-axis of the 3D NS recovery plots shows
\emph{gradient evaluations} rather than outer iterations:
Adam costs one value-and-gradient call per step, while L-BFGS's line
search can evaluate several trial points per step, so plotting per
iteration would understate L-BFGS's true forward-and-backward cost. The
x-axis includes every evaluation the optimizer made.
Per-task settings:
\begin{itemize}\itemsep0pt
  \item \textbf{H (conductivity recovery)}: L-BFGS, 1\,000 iters, on a
        sigmoid reparametrization that keeps the density in $[10^{-3},1]$ and
        makes the SIMP conductivity affine in the latent variable.
  \item \textbf{S (topology optimization)}: Adam $\eta{=}5{\times}10^{-2}$,
        2\,500 iters, patience 100, with a soft volume-fraction penalty
        of weight 50; L-BFGS, 100 iters, with the same penalty and box
        constraints (fails to reach Adam's optimum, see below);
        MMA via NLopt~\citep{svanberg1987mma}, at most 200 objective
        evaluations, stopping after 30 evaluations without a lower compliance,
        with a hard volume inequality and native box constraints.
  \item \textbf{F2 (drag minimization, Re=20)}: Adam $\eta{=}10^{-2}$, at
        most 100 iters, stopping once the loss has not dropped by 0.1\% in 10
        iterations.
  \item \textbf{F3 (3D NS IC recovery)}: Adam $\eta{=}10^{-3}$, 500 iters,
        patience 50; L-BFGS, 100 iters.
        L-BFGS+proj and Adam+proj variants apply a solenoidal projection to
        the gradient at each step (see next paragraph) with otherwise
        identical settings.
\end{itemize}

\paragraph{Pass criteria}
The \textit{Conv.} column of \cref{tab:merged} reports the outcome of the
benchmark's automated checks for the Adam runs on F2 and S, the L-BFGS run on H, and the
L-BFGS+proj run on F3. A run passes if its final objective is at most half the
initial one. For drag minimization, the final drag must also be at most 0.7
of the initial drag and lie within 30\% of the peer median. These thresholds
are loose enough that only broken gradients fail them, so they do not
distinguish between working solvers. \Cref{sec:optimization} compares the final objectives directly.

\paragraph{Solenoidal gradient projection (F3 variants)}
The L-BFGS+proj and Adam+proj variants on F3 apply a spectral
Helmholtz projection to the gradient \emph{before} each optimizer
update,
$\hat g^{\mathrm{df}}_i = \hat g_i - k_i (k\!\cdot\!\hat g) / |k|^2$,
where $\hat g$ is the FFT of the gradient on the periodic box. The
projection is implemented once in the harness and applied to all seven
fluid solvers.

\begin{figure}
  \centering
  \includegraphics[width=\linewidth]{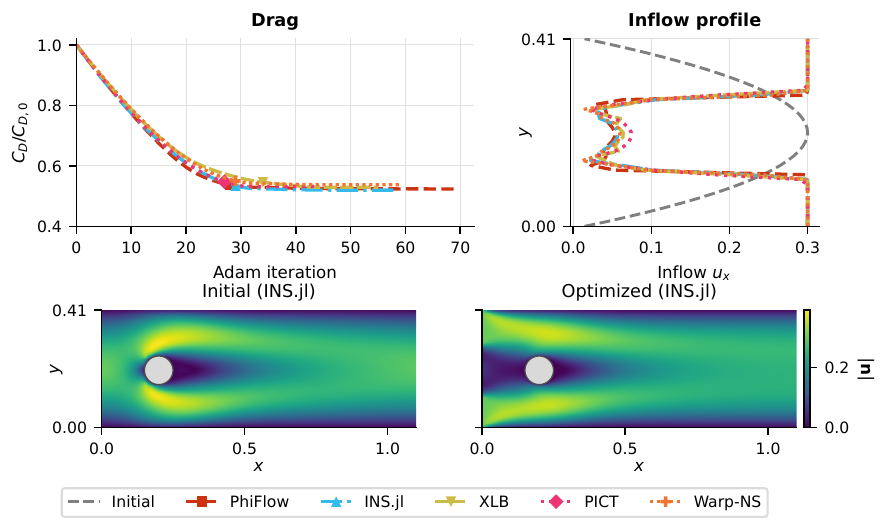}
  \caption{2D drag minimization on the DFG 2D-1 channel at Re\,=\,20.
    Top left: $C_d/C_{d,0}$ per Adam iteration, with a dot where the loss
    levels off. Top right: initial and optimized inlet profiles. Bottom:
    velocity magnitude near the cylinder before and after optimization for
    the solver with the largest drag reduction.}
  \label{fig:app:dragopt}
\end{figure}

\paragraph{2D drag minimization}
\Cref{fig:app:dragopt} shows the drag minimization (setup in \cref{app:effort}).
By $t=4$, the drag coefficient $C_d$ is within 0.1\% of its steady value at this
resolution. From initial drag coefficients of 5.3--5.9, the optimization reduces drag
by 46--48\% in 44--69 iterations with each of the five solvers that run the
channel (INS.jl 48.0\%, PhiFlow 47.7\%, XLB 47.1\%, PICT 46.8\%, Warp-NS
46.3\%). The loss levels off after 27--34 iterations.

\begin{figure}
  \centering
  \includegraphics[width=\linewidth]{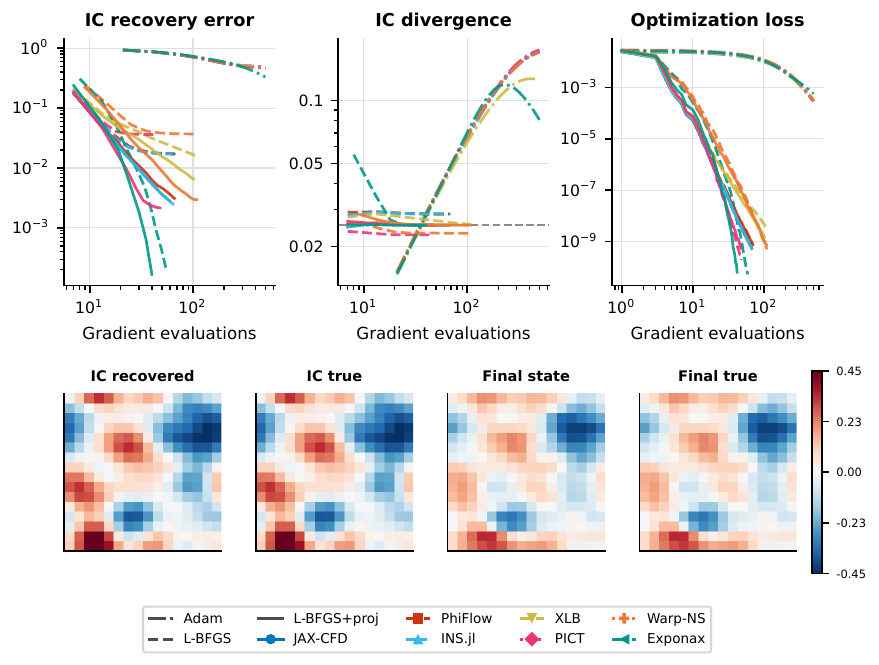}
  \caption{3D NS IC recovery overview for Adam, L-BFGS, and L-BFGS+proj.
    \emph{Top:} normalized IC error, IC divergence, and optimization loss vs.\ gradient evaluations across all solvers (color = solver, line style = optimizer).
    The dashed line marks the true IC divergence level.
    \emph{Bottom:} $u_x$ slice at convergence (L-BFGS+proj / PhiFlow): recovered IC, true IC, recovered final state, and true final state.}
  \label{fig:app:recovery}
\end{figure}

\paragraph{3D NS IC recovery}
\Cref{fig:app:recovery} compares three optimizer configurations on the 3D NS IC recovery task (setup in \cref{app:effort}).
We report the \emph{IC error} as the relative $\ell_2$ distance between the recovered and true initial velocity fields,
$\|\mathbf{u}^{*}_0 - \mathbf{u}_0\|_2 / \|\mathbf{u}_0\|_2$,
where $\mathbf{u}_0$ is the ground-truth IC and $\mathbf{u}^{*}_0$ is the optimizer iterate at the reported checkpoint.
Solver color and marker encode the backend; line style encodes the optimizer.
After its 500 iterations, Adam still has 38--49\% normalized IC error (Exponax 38\%, the other solvers 47--49\%), while L-BFGS reaches 0.01--4\% in 58--109 gradient evaluations.
Projecting each L-BFGS gradient onto divergence-free fields (L-BFGS+proj) lowers the final IC error to 0.01--0.85\% (below 0.5\% on every solver except XLB) and keeps the recovered IC divergence at the level of the true IC throughout.
Projecting Adam's gradient as well (Adam+proj) does not lower the IC error on any solver: the final error changes by at most 1.1 percentage points on five solvers and rises by 8 on XLB and 17 on Exponax.
The bottom row shows $u_x$ at the middle $z$-slice for a representative solver (L-BFGS+proj/PhiFlow).

\begin{figure}
  \centering
  \includegraphics[width=\linewidth]{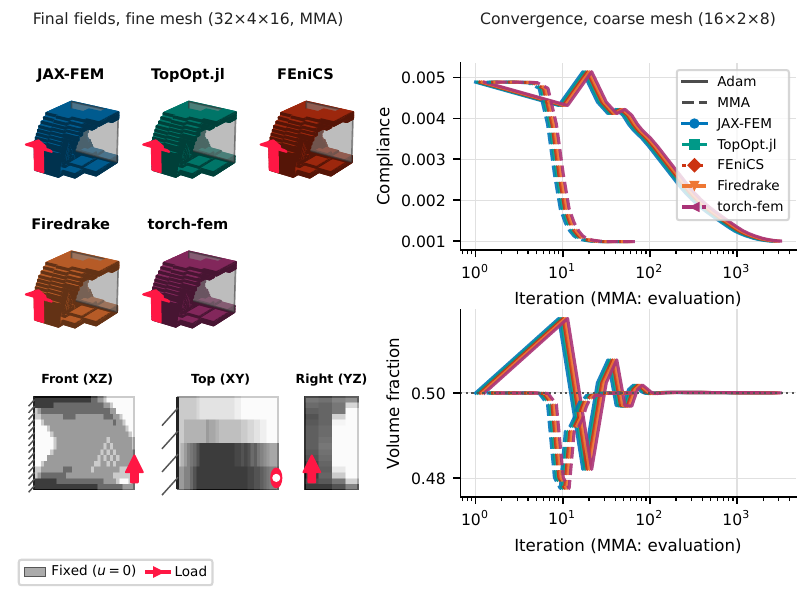}
  \caption{Structural topology optimization overview.
    \emph{Left:} final optimized density fields from a finer-mesh ($32{\times}4{\times}16$) MMA run, both as 3D voxel renderings (one per solver) and projected along each principal axis.
    \emph{Right:} compliance and volume-fraction vs.\ iteration for Adam and MMA (line style = optimizer, color = solver) on the $16{\times}2{\times}8$ mesh of the benchmark task. The MMA x-axis counts objective evaluations: MMA stops after 35--40, Adam runs its full budget of 2\,500 iterations.}
  \label{fig:app:topopt}
\end{figure}

\paragraph{Topology optimization}
\Cref{fig:app:topopt} summarizes the structural topology optimization task ($[0,2]\times[0,1]\times[0,1]$ cantilever beam, $16\times2\times8$ mesh, SIMP with $p=3$, Adam).
All solvers converge to the same final compliance (${\approx}0.001$).
The density fields of a finer-mesh MMA run (left) are visually identical across solvers.

We also evaluated L-BFGS and MMA~\citep{svanberg1987mma} on this task.
After 100 quasi-Newton steps, L-BFGS final compliance ranged from $0.00108$ (JAX-FEM) to $0.00112$ (FEniCS, Firedrake, TopOpt.jl), 7--12\% above the Adam solution ($0.00100$).
MMA reaches compliance ${\approx}0.00099$ on every solver within 35--40 objective evaluations, slightly below Adam's solution in under 2\% of Adam's iteration budget.

\begin{figure}
  \centering
  \includegraphics[width=\linewidth]{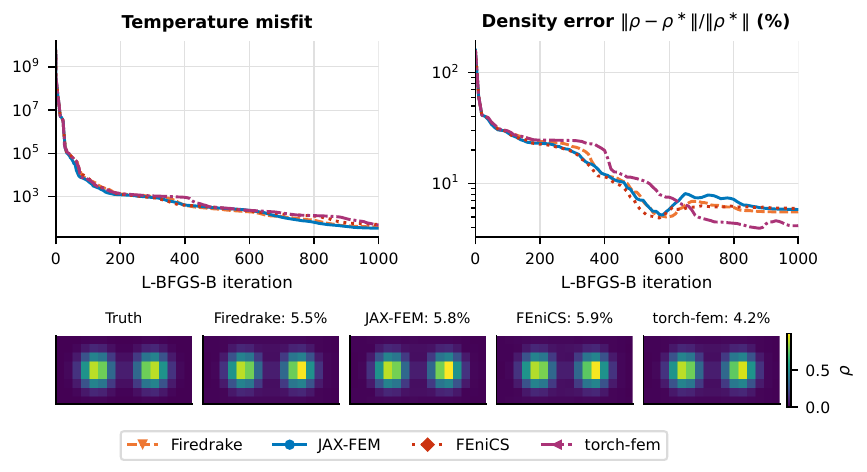}
  \caption{Thermal conductivity recovery with L-BFGS-B. \emph{Top:} temperature misfit and relative error of the recovered SIMP density against iteration, per solver. \emph{Bottom:} ground-truth and recovered density fields.}
  \label{fig:app:conductivity_recovery}
\end{figure}

\paragraph{Conductivity recovery}
\Cref{fig:app:conductivity_recovery} shows convergence for the thermal conductivity inversion task.
The ground-truth conductivity field has two Gaussian peaks. We recover it from the steady-state temperature distribution under a known heat source by minimizing the $\ell_2$ error between simulated and observed temperatures.
Starting from a uniform field, L-BFGS-B lowers the misfit of all four solvers from $6.1\times10^{9}$ to 34--51 within 1\,000 iterations (JAX-FEM 34, Firedrake 39, torch-fem 49, FEniCS 51), leveling off after 735--846 iterations.
The recovered SIMP density is 4.2--5.9\% from the truth in relative $\ell_2$ norm, and the four recovered fields differ from each other by 1--3.5\% of the truth's norm.
We attribute the spread to the poor conditioning of the inversion: many conductivity fields fit the temperatures almost equally well, and which one the optimizer reaches depends on its path.

\subsection{Forced Kolmogorov flow}
\label{app:kolmogorov}

\paragraph{Setup}
The forced F2 setup is 2D turbulence on the periodic box $[0,2\pi)^2$,
$\partial_t\mathbf{u} + (\mathbf{u}\cdot\nabla)\mathbf{u} = -\nabla p
+ \nu\nabla^2\mathbf{u} + \mathbf{f} - \alpha\mathbf{u}$,
$\nabla\cdot\mathbf{u} = 0$, with the body force
$\mathbf{f} = (A\sin\kappa y, 0)$, $A = 1$, $\kappa = 4$, linear drag
$\alpha = 0.1$ and $\nu = 2\times10^{-3}$.
The linear drag is needed because in 2D turbulence, energy cascades to the
largest scales. With it, the flow settles into a chaotic, statistically steady
state, which a $64^2$ grid resolves only roughly. Every 0.1 time units, the
$64^2$ solution departs by about 2\% from a $4\times$ finer run.

\paragraph{Forcing inputs}
On periodic 2D boxes, the F2 interface accepts a cell-centered force field and a
drag coefficient as optional inputs, which are not differentiated. Each solver
applies them inside its own time step, in both the forward pass and the VJP, and
maps the force to its native grid locations the same way it maps the initial
condition (\cref{tab:app:forcing}). With a force, XLB sizes its lattice Mach
number from a bound that does not depend on the solution, so a flow spun up from
rest stays at the target Mach number.

\begin{table}[h]
  \centering\small
  \caption{How each solver applies the Kolmogorov force and the linear drag.}
  \label{tab:app:forcing}
  \begin{tabular}{@{}lll@{}}
    \toprule
    \textbf{Solver} & \textbf{Force} & \textbf{Drag} \\
    \midrule
    JAX-CFD & forcing callable, averaged onto the faces & explicit, every RK4 stage \\
    PhiFlow & resampled to the faces, every RK4 stage & every RK4 stage \\
    INS.jl & projected staggered force in the right-hand side & every RK4 stage \\
    XLB & exact-difference method after collision & semi-implicit, half-force velocity \\
    PICT & velocity source, set every step & explicit, start of step \\
    Warp-NS & tentative-velocity kernel & explicit \\
    Exponax & spectral curl of the force, as its injection term & its native drag term \\
    OpenFOAM & \multicolumn{2}{l}{not supported: icoFoam has no momentum source term} \\
    \bottomrule
  \end{tabular}
\end{table}

\paragraph{Ground truth}
The ground truth is an Exponax run at $1024^2$ (ETDRK2 with 2/3-rule dealiasing,
CFL number about 0.1) from three test states, each spun up from a random field
for 100 time units at $512^2$. A $512^2$ run from the same state differs from it by
0.3\% after one time unit and by 2.3--3.2\% after four, at a vorticity
correlation of at least 0.997. It stays above 0.95 correlation for 6.5 to 7.5
time units, beyond the scored horizon of four. As an independent check, JAX-CFD
converges to the truth at observed orders 1.6 and 1.9 between $64^2$ and
$256^2$, down to the 0.3\% floor at $512^2$.

\paragraph{Checks}
\textit{Laminar forcing} (\texttt{forward/kolmogorov\_laminar}). From rest at
$\nu = 0.05$, the flow stays a shear flow whose advection term vanishes, so
$u = (A/\lambda)(1 - e^{-\lambda t})\sin(\kappa y)$, $v = 0$ with
$\lambda = \nu\kappa^2 + \alpha$ is exact. Solvers run $N \in \{32, 64, 128\}$
to $t = 2$ with a time step small enough that the error is spatial. A solver
passes with at most 5\% error at $N = 64$ and an observed order of at least
0.8. With a correct second-order scheme, the error at $N = 64$ is about 1\%,
and with a convention error (a missing $2\pi/L$, a flipped sign, a force half a
cell off), it is 20\% or more.

\textit{Turbulent forward accuracy} (\texttt{forward/kolmogorov}). Each solver
runs grids from $32^2$ to $256^2$ at the CFL number of $64^2$ with
$\Delta t = 0.01$, from the three test states to $t = 1, 2, 3, 4$, and is scored
on $64^2$ against the truth. It passes if its observed order from $64^2$ to
$256^2$ at $t = 1$ is at least 0.8 and none of its runs from $64^2$ up diverges. The bound
follows from the scheme's consistency and does not depend on the other solvers.

\textit{Chained gradients} (\texttt{gradient/chain\_fd\_check}). In a training
loss, each interval's error is normalized by the target's norm, so each call in
a chain receives a far smaller cotangent than in the single-call checks. The
chain consists of 16 calls of 0.1 time units at $64^2$, starting from a
test state plus a 1\% divergence-free offset. Its loss is the mean relative
squared distance to the solver's own unperturbed chain. The VJP is compared with
central differences along 8 divergence-free directions at
$\varepsilon \in \{1, 0.1, 0.01\}$. A solver passes with a cosine of at least
0.99, a relative error of at most $10^{-2}$, and an error no more than 50 times
the peer median.

\begin{table}[h]
  \centering\small
  \caption{Kolmogorov-flow checks. \emph{Laminar}: relative error against the
    exact solution at $t = 2$. \emph{Turbulent}: median relative error against
    the ground truth at $t = 1$ on each grid, and the observed order from $64^2$
    to $256^2$. \emph{Chained}: median relative error and cosine of the VJP at
    the best $\varepsilon$, with PICT's cotangent scaling (see below).}
  \label{tab:app:kolmogorov}
  \begin{tabular}{@{}lccc@{}}
    \toprule
    & \multicolumn{3}{c}{\textbf{Laminar error}} \\
    \cmidrule(l){2-4}
    \textbf{Solver} & $N=32$ & $64$ & $128$ \\
    \midrule
    Exponax & 3.0e-6 & 3.5e-5 & 1.3e-5 \\
    JAX-CFD, INS.jl, PhiFlow & 2.9e-2 & 7.4e-3 & 1.8e-3 \\
    PICT & 2.9e-2 & 7.1e-3 & 1.7e-3 \\
    Warp-NS & 3.0e-2 & 7.7e-3 & 2.2e-3 \\
    XLB & 7.8e-2 & 1.9e-2 & 4.8e-3 \\
    \bottomrule
  \end{tabular}

  \vspace{6pt}
  \begin{tabular}{@{}lcccccc@{}}
    \toprule
    & \multicolumn{5}{c}{\textbf{Turbulent error}} & \multicolumn{1}{c}{\textbf{Chained}} \\
    \cmidrule(lr){2-6}\cmidrule(l){7-7}
    \textbf{Solver} & $32^2$ & $64^2$ & $128^2$ & $256^2$ & order & error / cosine \\
    \midrule
    Exponax & 0.168 & 0.039 & 0.0046 & 0.0028 & 1.89 & 6.4e-4 / 1.000 \\
    JAX-CFD & 0.226 & 0.086 & 0.029 & 0.0077 & 1.73 & 2.1e-4 / 1.000 \\
    INS.jl & 0.226 & 0.086 & 0.029 & 0.0077 & 1.73 & 2.6e-4 / 1.000 \\
    XLB & 0.135 & 0.032 & 0.0063 & 0.0032 & 1.66 & 1.2e-4 / 1.000 \\
    PICT & 0.258 & 0.092 & 0.034 & 0.012 & 1.46 & 1.8e-4 / 1.000 \\
    PhiFlow & 0.206 & 0.082 & 0.028 & 0.0076 & 1.72 & 5.3e-5 / 1.000 \\
    Warp-NS & \multicolumn{5}{c}{non-finite on every grid by $t = 3$} & excluded \\
    \bottomrule
  \end{tabular}
\end{table}

\paragraph{Results}
\Cref{tab:app:kolmogorov} lists the results. Every solver passes the laminar
check. Exponax represents the single forced mode exactly, so its laminar error
is float32 round-off. Every solver that runs the turbulent flow stably passes
both remaining checks. On the chain, each solver's finite differences agree
with its VJP to within 2\% (median over directions) at every $\varepsilon$.
JAX-CFD and INS.jl advance this flow identically up to float32 round-off. Over
the 40 intervals scored in \cref{app:sil}, their uncorrected kinetic energies
differ by at most $6\times10^{-5}$ in relative terms, so their forward-check
errors and their test errors in \cref{tab:app:sil} coincide to the printed
digits. PhiFlow's default periodic pressure
solve adds a Lagrange multiplier, which makes the system indefinite, and some of
its CG solves then stop unconverged without an error. With the default, two of
the three $256^2$ runs turned non-finite and the chained gradient was off by
12\%. Mosaic therefore runs CG on the singular system directly, and PhiFlow
passes with that setting.

Warp-NS integrates with explicit Euler and central differences, which is
linearly unstable on this flow unless $\Delta t \lesssim 4\nu/u^2$, a bound that
does not shrink with the grid. Its turbulent runs turn non-finite on every grid,
which the forward check records as an explained anomaly. It is excluded from
the chained check, where the unstable modes dominate both the VJP and the finite
differences. OpenFOAM's icoFoam has no momentum source term, so OpenFOAM runs
neither forward check, and it has no VJP (\cref{app:exclusions}).

\paragraph{PICT's adjoint tolerance}
PICT's linear solves stop once the RMS of their residual falls below an absolute
tolerance, $10^{-5}$ in float32 by default, and its adjoint solves use the same
tolerance. When the cotangent is small, the adjoint residual starts near that
tolerance and the solves stop early, so the VJP is no longer linear in the
cotangent. Single calls with $O(1)$ cotangents are unaffected. On the 16-call
chain, with cotangents of about $10^{-4}$ per entry, the gradient with respect to
a state 1, 3, 8, and 16 calls before the loss disagreed with finite differences by
0.04\%, 3.7\%, 10\%, and 73\%, and the chained check reported a median error of
0.89 at a cosine of 0.51. The wrapper therefore scales each cotangent to a
largest entry of one before PICT's backward pass and scales the gradient back,
which is exact for a linear VJP and makes the tolerance relative to the
cotangent. All results in this paper use this wrapper.

\subsection{Solver-in-the-loop study}
\label{app:sil}

\paragraph{Corrector}
A periodic CNN with three $5\times5$ convolutional layers and 8 hidden channels
reads the velocity and the phase of the forcing ($\sin\kappa y$,
$\cos\kappa y$), since the force breaks the translation symmetry in $y$. It
returns a stream function $\psi$, and the correction is its spectral curl, which
is exactly divergence-free with zero mean. The readout weights are initialized
to zero, so the correction is zero before training. Input and output scales are
fixed by the forcing rather than by any solver's data, and the corrector has the
same architecture and initialization for every solver.

\paragraph{Training}
Each solver runs the coarse configuration of \cref{app:kolmogorov} ($64^2$,
$\Delta t = 0.01$) and is corrected every 0.1 time units. Its targets are its
own $256^2$ run, restricted to $64^2$, from four training states and one
validation state, spun up like the test states of \cref{app:kolmogorov}.
The three trained arms are \emph{supervised}, on fixed pairs (the solver's output
from reference frame $k$, and reference frame $k+1$) with no solver in the
training graph; \emph{stopped}, on a corrected 16-interval rollout with the
gradient stopped at every solver output; and \emph{SIL}, on the same rollout
differentiated through the solver. All arms are trained on the same windows
with Adam, learning rate $10^{-3}$, and global-norm clipping for 256 updates,
for each of three network seeds. These settings are shared by all arms and solvers and
were not tuned per arm, and the rollout length is fixed at 16 intervals from
the first update rather than increased over a curriculum. The loss is the mean over the rollout of
$\|s - t\|^2 / \|t\|^2$, without per-solver normalization. We select the
checkpoint by validation on the separate state every 32 updates, with update 0,
the uncorrected solver, as a candidate. Every selected and final corrector is then
evaluated against the solver's own reference on the three test states for 40
intervals, and rolled out for 200 intervals to check stability.

\paragraph{Gates and solvers}
A solver is admitted if its $256^2$ reference differs from a $512^2$ run by at
most a quarter of the gap between its coarse run and the reference, if chained
calls reproduce one long call, and if the gradient of the training loss through
16 intervals matches central finite differences to within 5\% at a zero and at
a random nonzero readout. We check these gates, and that the losses stay finite
in every arm, in a short version of the study (4-interval rollouts, 32 updates,
one seed). JAX-CFD, Exponax, XLB, INS.jl, and PICT pass all of them, with reference
ratios of 0.003 to 0.093 and gradient errors of $5.8\times10^{-5}$ to
$1.8\times10^{-3}$. PICT runs with the cotangent scaling of
\cref{app:kolmogorov}, which fixes the one defect found in the study. Warp-NS
diverges on the flow (\cref{app:kolmogorov}).
PhiFlow is excluded because its $256^2$ run, which the corrector would train
against, turns non-finite within the study's horizon from some of the
reference states. OpenFOAM is forward-only.
We run the full study only for solvers that complete it in about an hour on the A100
(29 to 67 minutes for the four that do). That excludes PICT, whose updates
through the solver take 12 times as long as JAX-CFD's in the short version and
whose full run takes most of a day.

\paragraph{Native solver state}
For staggered and lattice solvers, a chain of calls that restarts from the
cell-centered velocity is a different map from one long call, because every
restart involves interpolating to the cell centers and back.
\Cref{tab:app:sil-closure} shows how far such chains drift. For the study,
JAX-CFD, INS.jl, and XLB take and return their native state (face velocities or
lattice populations) as an optional input and output, which makes chains exact
up to round-off. PICT and Exponax are collocated and carry no history between
steps, so they restart exactly without it.

\begin{table}[h]
  \centering\small
  \caption{Closure of chained calls: median relative $\ell_2$ distance over the
    three test states between a chain of one-interval calls and one call over
    the same time.}
  \label{tab:app:sil-closure}
  \begin{tabular}{@{}lcccccc@{}}
    \toprule
    & \multicolumn{3}{c}{\textbf{Restart from velocity}} & \multicolumn{3}{c}{\textbf{Native state}} \\
    \cmidrule(lr){2-4}\cmidrule(l){5-7}
    \textbf{Solver} & 2 calls & 16 & 40 & 2 calls & 16 & 40 \\
    \midrule
    JAX-CFD & 1.0e-2 & 0.13 & 0.55 & 0 & 0 & 0 \\
    INS.jl & 1.0e-2 & 0.13 & 0.55 & 0 & 0 & 0 \\
    XLB & 5.3e-3 & 2.3e-2 & 9.7e-2 & 1e-6 & 1e-6 & 3e-6 \\
    PICT & 0 & 0 & 0 & $-$ & $-$ & $-$ \\
    Exponax & 1e-6 & 7e-6 & 5e-5 & $-$ & $-$ & $-$ \\
    \bottomrule
  \end{tabular}
\end{table}

\begin{table}[h]
  \centering\small
  \caption{Solver-in-the-loop results. Mean relative test error over 40
    intervals for each of the three network seeds, the uncorrected solver's
    error, and the wall time per training update at a 16-interval rollout.}
  \label{tab:app:sil}
  \setlength{\tabcolsep}{4pt}
  \begin{tabular}{@{}lccccc@{}}
    \toprule
    & \multicolumn{3}{c}{\textbf{Test error per seed}} & & \textbf{s / update} \\
    \cmidrule(lr){2-4}\cmidrule(l){6-6}
    \textbf{Solver} & Supervised & Stopped & SIL & Uncorrected & stopped / SIL \\
    \midrule
    JAX-CFD & 0.117 0.100 0.108 & 0.151 0.161 0.167 & 0.193 0.151 0.189 & 0.243 & 0.79 / 1.02 \\
    INS.jl & 0.117 0.100 0.108 & 0.151 0.161 0.167 & 0.193 0.151 0.189 & 0.243 & 1.22 / 2.84 \\
    Exponax & 0.068 0.075 0.067 & 0.063 0.061 0.065 & 0.076 0.068 0.063 & 0.091 & 0.80 / 1.04 \\
    XLB & 0.069 0.061 0.063 & 0.078 0.085 0.068 & 0.084 0.073 0.102 & 0.086 & 0.85 / 1.39 \\
    \bottomrule
  \end{tabular}
\end{table}

\paragraph{Results}
All 36 trainings complete their 256 updates. The SIL arm's gradient norms are
10 to 20 times those of the stopped arm but stay bounded. None of the 216 long
rollouts turns non-finite, and no corrector from any arm is less stable than
the uncorrected solver. The peak kinetic energy of a corrected rollout reaches 1.06
to 1.50 times the uncorrected mean, against 1.09 to 1.45 for the uncorrected
solver itself. Averaged over the seeds (\cref{tab:app:sil}), supervised training
gives the lowest error on every solver except Exponax, where the stopped arm is
lower, and SIL's average is never better than the stopped arm's. Once per
solver, an individual SIL seed does end below it, as XLB's second seed does
(0.073 against 0.085). On XLB, SIL's mean matches the uncorrected error (0.086), since its third
seed ends above it. Updates through the solver take 0.79 to 2.84\,s, whereas
supervised updates take about 0.02\,s, because the solver's outputs for the pairs
are computed once in advance.

\paragraph{Reproduction}
The study is not one of the benchmark's suites, but its code (the native
solver state, the full and short training experiments, and the scripts that
produce these tables) is part of the release tag and reproduction package
(\cref{app:software}), with instructions for running it in
\texttt{studies/kolmogorov-sil/}.

\section{Solver documentation}
\label{app:solvers}

This appendix documents all solvers evaluated for inclusion in Mosaic,
covering implementation details, wrapping approach, and known limitations.
Solvers included in the benchmark appear in Table~\ref{tab:solvers}.
Excluded solvers and the reasons for their exclusion are listed in
\cref{app:excluded}.

\subsection{Included solvers}
\label{app:included}

\paragraph{JAX-CFD~\cite{kochkov2021machine}}
JAX-native incompressible flow solver on a staggered MAC grid with
finite-difference advection and spectral (FFT) pressure projection.
Gradients via JAX source-transformation AD.
Runs the periodic setups of F2 and F3 but not the DFG channel
(\cref{app:exclusions}).

\paragraph{PhiFlow~\cite{holl2024phiflow}}
Pressure projection with central differences and RK4 on the periodic boxes,
and semi-Lagrangian advection in the channel.
Supports PyTorch, JAX, and TensorFlow backends, enabling gradient
computation through the same simulation code across AD frameworks.
Mosaic runs it on JAX, with each CG pressure solve differentiated by an
adjoint CG solve.

\paragraph{INS.jl~\cite{agdestein2024ins}}
IncompressibleNavierStokes.jl: Julia staggered finite-volume pressure-projection
solver, run on the GPU through its CUDA backend. In the channel, the cylinder is
represented by Brinkman penalization. Differentiates through the time loop,
including with respect to $\nu$, via Zygote.jl reverse-mode AD.

\paragraph{XLB~\cite{ataei2024xlb}}
JAX-native lattice Boltzmann solver supporting D2Q9 (2D) and D3Q27 (3D) stencils on
GPU, run with BGK collision in float64. Gradients via source-transformation AD
through the collision-streaming loop, checkpointed in nested blocks of about
$\sqrt{T}$ steps.
The BGK lattice Boltzmann method~\cite{qian1992lattice} recovers the
incompressible Navier--Stokes equations only to $O(\mathrm{Ma}^2)$. At fixed
$\Delta t$, the Mach number $\mathrm{Ma} = u\,\Delta t / (c_s\,\Delta x)$, with
$c_s=1/\sqrt{3}$, grows with grid refinement, so the compressibility error would
not decrease with spatial resolution. The wrapper therefore runs each prescribed
step as $k=\lceil \mathrm{Ma}/\mathrm{Ma}_\text{target}\rceil$ lattice steps of
$\Delta t/k$, with $\mathrm{Ma}_\text{target}=0.1$, lowered in proportion to $1/N$
above $N=64$ on the periodic boxes.

\paragraph{PICT~\cite{franz2025pict}}
GPU-accelerated differentiable incompressible Navier--Stokes solver built on
PyTorch with custom CUDA kernels implementing the PISO algorithm. Supports
reverse-mode AD through the full time loop, including with respect to $\nu$,
and handles multi-block curvilinear grids. The Mosaic Tesseract runs the
periodic boxes on a single block and the DFG channel on a body-fitted mesh of
eight Cartesian blocks and four O-grid blocks around the cylinder, with an
advective outflow.

\paragraph{Warp-NS~\cite{macklin2024warp}}
Custom incompressible NS solver built with NVIDIA Warp CUDA kernels,
implementing Chorin projection with explicit Euler time stepping. On the
periodic boxes, it solves the pressure Poisson equation with an FFT, and in the
channel, it uses a staggered grid, a staircase cylinder, and a preconditioned CG
pressure solve.
Gradients via the \texttt{wp.Tape} kernel-level VJP, with a hand-written
adjoint for each channel CG solve.
The Warp framework provides differentiable kernels but no ready-made
incompressible flow solver, so the FD stencils, pressure Poisson solves,
and time integration were implemented from scratch using Warp primitives.
Warp-NS is a best-effort implementation by the Mosaic team and an example
of the integration cost when a kernel toolkit provides no built-in solver.

\paragraph{Exponax~\cite{koehler2024apebench}}
Spectral incompressible Navier--Stokes solver supporting both 2D
(streamfunction-vorticity) and 3D (Leray-projected velocity) formulations,
integrated with an exponential time-differencing Runge-Kutta (ETDRK) scheme.
Incompressibility is enforced to machine precision by construction.
Gradients via JAX source-transformation AD.
Used in Mosaic for the periodic setups of F2 and F3, as the converged
reference for F3, and as the ground truth of the Kolmogorov-flow checks
(\cref{app:kolmogorov}).

\paragraph{FEniCS~\cite{alnaes2015fenics}}
Finite-element solver using Q1 elements (CG1 on hexahedra) for thermal and structural problems.
dolfin-adjoint~\cite{farrell2013automated} automates the discrete adjoint by replaying the forward tape.

\paragraph{Firedrake~\cite{rathgeber2016firedrake}}
Mirrors the FEniCS Q1/CG1 formulation for structural and thermal problems.
Differentiates via firedrake-adjoint~\cite{mitusch2019dolfinadjoint}, providing an independent
tape-based adjoint implementation for cross-validation.

\paragraph{JAX-FEM~\cite{xue2023jaxfem}}
Solves heat conduction and linear elasticity with trilinear HEX8 finite
elements in JAX on the GPU, with Jacobi-preconditioned BiCGSTAB for the forward
and adjoint systems. Gradients via JAX AD through the assembly, with the linear
solve differentiated by an adjoint solve.

\paragraph{TopOpt.jl~\cite{huang2021topoptjl}}
SIMP topology optimization for linear elasticity with HEX8 elements in
Julia, using analytical adjoint sensitivities.

\paragraph{torch-fem~\cite{meyer2024torchfem}}
PyTorch finite-element solver for heat conduction and linear elasticity with
linear HEX8 elements, with Jacobi-preconditioned CG for the forward and adjoint
solves. Gradients via PyTorch autograd through the assembly, with the linear
solve differentiated by an adjoint solve.
GPU-accelerated via PyTorch sparse operations.

\paragraph{OpenFOAM~\cite{weller1998tensorial} (reference)}
Runs the icoFoam incompressible PISO solver on the periodic boxes and
\texttt{foamRun} with the incompressibleFluid solver (PIMPLE) on a body-fitted
O-grid mesh in the DFG channel, as a forward-only reference
baseline. No reverse-mode AD available.

\paragraph{deal.II~\cite{bangerth2007dealii,arndt2023dealii} (reference)}
Solves thermal and structural problems with Q1 elements using the
industry-grade C++ finite-element library.
No native reverse-mode AD is available; used as a forward-only reference
baseline for S and H, analogously to OpenFOAM for the fluid domains.

\subsection{Excluded solvers}
\label{app:excluded}

We document solvers that were evaluated for inclusion but ultimately
excluded, along with the specific reason in each case. Practitioners
looking for a particular solver can check whether it was considered and
why it was left out.

\paragraph{WaterLily.jl~\cite{waterlily2024}}
Forward-mode AD only, via \texttt{ForwardDiff.jl} dual numbers. No
Zygote reverse-mode support is documented. A VJP can be emulated by
contracting $N$ forward-mode JVPs, but the cost scales as
$O(N^d \cdot T_{\mathrm{fwd}})$, matching central finite differences.

\paragraph{JAX-Fluids~\cite{jaxfluids2022}}
Targets compressible flow. Recovering the incompressible limit at low Mach
number ($\mathrm{Ma} \approx 0.01$) requires acoustic sub-stepping whose
iteration count depends on Ma, grid size, and time step simultaneously.

\paragraph{Commercial solvers (ANSYS, COMSOL, Abaqus)}
Excluded for reproducibility: closed-source licenses prevent redistribution
of solver binaries or independent verification of gradient implementations.

\paragraph{FEniCS-NS~\cite{alnaes2015fenics}}
The FEniCS incompressible NS solver (CG1/CG1 elements, IPCS time-stepping)
was evaluated for fluid benchmarks but excluded for two independent reasons.
Without SUPG/GLS advection stabilization, stability requires
$\mathrm{Pe}_h = Uh/(2\nu) \lesssim 1$, so at $\nu < 0.01$ on the $N=64$
benchmark grid its errors were orders of magnitude above the other solvers'.
Independently, its serial wall-clock time grew faster than quadratically in
$N$, and a single VJP at $N=64$ exceeded the per-trial limit of the cost
sweeps.
FEniCS remains included for structural and heat-transfer domains (H, S),
where neither constraint applies.

\section{Excluded tasks and what would lift each exclusion}
\label{app:exclusions}

Setup compatibility is one of the protocol's metrics, so every task a
solver does not run is recorded in the repository with a category and a
reason. \Cref{tab:app:exclusions} lists the exclusions that affect the
tasks and checks of this paper, with the change that would lift each.
Categorical exclusions follow from the solver's method or from a missing
adjoint, and the solver cannot be compared without a different
discretization or an adjoint implementation. A contributor could lift the
others. Flagged results with a recorded explanation are kept and not
listed. Solvers left out of the
benchmark altogether are described in \cref{app:excluded}.

\begin{table}[h]
  \centering\small
  \caption{Solvers excluded from tasks or checks in this paper. Categories
    are those recorded in the repository, except PICT's, which is the
    runtime criterion of the solver-in-the-loop study (\cref{app:sil}).}
  \label{tab:app:exclusions}
  \renewcommand{\arraystretch}{1.15}
  \begin{tabularx}{\linewidth}{@{}>{\raggedright}p{1.75cm}>{\raggedright}p{2.15cm}>{\raggedright}p{1.7cm}>{\raggedright}X>{\raggedright\arraybackslash}X@{}}
    \toprule
    \textbf{Solver} & \textbf{Tasks} & \textbf{Category} & \textbf{Reason}
      & \textbf{What would lift it} \\
    \midrule
    Exponax & F2 channel: cylinder check, drag optimization & Categorical
      & A Fourier pseudo-spectral method represents only periodic fields, so
        it cannot express walls, an inflow, or an outflow.
      & Nothing within the method. The channel needs a different
        discretization. \\
    JAX-CFD & F2 channel: cylinder check, drag optimization & Not implemented
      & Version 0.2.1 supports no inflow in its advection, no zero-gradient
        outlet for face-centered velocity, only scalar boundary values, and
        no obstacles.
      & Upstream support for these boundaries, or hand-written convection,
        boundary conditions, and obstacle on its grid primitives. \\
    OpenFOAM & All gradient-based experiments (F2, F3) & Categorical
      & The standard solvers are forward-only C++ without an AD path.
      & An adjoint-capable variant, such as DAFoam or OpenFOAM-AD, wrapped as
        a separate backend. \\
    OpenFOAM & F2 Kolmogorov flow: laminar and turbulent forward checks
      & Not implemented
      & The wrapper runs icoFoam, which has no momentum source term, so it
        cannot apply the body force or the drag.
      & Running the \texttt{foamRun} \texttt{incompressibleFluid} solver, which
        applies both through an \texttt{fvModel}. \\
    deal.II & All gradient-based experiments (H, S) & Categorical
      & The library offers no reverse-mode VJP through its solves, and the
        wrapper is forward-only.
      & A hand-written discrete adjoint in the wrapper, one transposed linear
        solve per gradient for these steady linear problems. \\
    Warp-NS & F2 Kolmogorov flow: chained gradient check; solver-in-the-loop
        study & Unstable
      & Explicit Euler with central advection is linearly unstable at the
        time steps of the forced Kolmogorov flow. It diverges on every grid,
        and over a chain the unstable modes dominate the VJP and the finite
        differences.
      & A stable time integrator or advection scheme. \\
    PhiFlow & Solver-in-the-loop study & Unstable
      & Its own finest-grid run, the training target, turns non-finite within
        the study's horizon from some reference states. The cause has not been
        established.
      & A diagnosis of the divergence. \\
    PICT & Full solver-in-the-loop study & Runtime
      & The full run would take most of a day, because PICT's linear solves
        are bound by device synchronization.
      & Fewer synchronizations per solve, or a larger compute budget. \\
    \bottomrule
  \end{tabularx}
\end{table}

\end{document}